\documentclass[10pt,journal,compsoc]{IEEEtran}

\ifCLASSOPTIONcompsoc

  \usepackage[nocompress]{cite}
\else

  \usepackage{cite}
\fi

\ifCLASSINFOpdf

\else

\fi

\usepackage[utf8]{inputenc} % allow utf-8 input
\usepackage[T1]{fontenc}    % use 8-bit T1 fonts
\usepackage{hyperref}       % hyperlinks
\usepackage{url}            % simple URL typesetting
\usepackage{booktabs}       % professional-quality tables
\usepackage{amsfonts}       % blackboard math symbols
\usepackage{nicefrac}       % compact symbols for 1/2, etc.
\usepackage{microtype}      % microtypography
\usepackage{lipsum}
\usepackage{fancyhdr}       % header
\usepackage{graphicx}       % graphics
\graphicspath{{media/}}     % organize your images and other figures under media/ folder
\usepackage{enumitem}
\usepackage{newclude}
\usepackage{threeparttable}
\usepackage{colortbl}
\usepackage{tabularray}
\usepackage{balance}
\usepackage{caption}
\usepackage{makecell}
\usepackage{multirow}

\usepackage{color}

\usepackage{subfiles}

\begin{document}

\title{Computational Protein Science in the Era of Large Language Models (LLMs)}

\author{Wenqi Fan, Yi Zhou, Shijie Wang, Yuyao Yan, Hui Liu, Qian Zhao, Le Song, and Qing Li

\IEEEcompsocitemizethanks{
\IEEEcompsocthanksitem W. Fan is with the Department of Computing and Department of Management and Marketing, The Hong Kong Polytechnic University. E-mail: \href{mailto:wenqifan03@gmail.com}{wenqifan03@gmail.com}.% <-this % stops an unwanted space
\IEEEcompsocthanksitem Y. Zhou, S. Wang, and Q. Li are with the Department of Computing, The Hong Kong Polytechnic University. Email: \{echo-yi.zhou, shijie.wang\}@connect.polyu.hk, \href{mailto:csqli@comp.polyu.edu.hk}{csqli@comp.polyu.edu.hk}.
\IEEEcompsocthanksitem Y. Yan is with the Department of Health Technology and Informatics, The Hong Kong Polytechnic University. E-mail: \href{mailto:yanyuyao69@gmail.com}{yanyuyao69@gmail.com}.
\IEEEcompsocthanksitem Hui Liu is with Michigan State University. Email: \href{mailto:liuhui7@msu.edu}{liuhui7@msu.edu}.
\IEEEcompsocthanksitem Qian Zhao is with the Department of Applied Biology and Chemical Technology, The Hong Kong Polytechnic University. Email: \href{mailto:q.zhao@polyu.edu.hk}{q.zhao@polyu.edu.hk}.
\IEEEcompsocthanksitem Le Song is with GenBio AI and Mohamed bin Zayed University of Artificial Intelligence. Email: \href{mailto:le.song@mbzuai.ac.ae}{le.song@mbzuai.ac.ae}.
}

\thanks{(Corresponding authors: Dr. Wenqi Fan and Prof. Qing Li.)}
}

\markboth{Journal of \LaTeX\ Class Files,~Vol.~14, No.~8, August~2015}%
{Shell \MakeLowercase{\textit{et al.}}: Bare Demo of IEEEtran.cls for Computer Society Journals}

\IEEEtitleabstractindextext{

\begin{abstract}

Proteins are macromolecules that play essential roles in almost all essential life activities, such as immunity, digestion, disease regulation, etc.
Considering the significance of proteins, computational protein science has always been a critical field of scientific research, dedicated to revealing knowledge and developing applications within the protein sequence-structure-function paradigm. 
In the last few decades, Artificial Intelligence (AI) has made a significant impact in computational protein science, leading to notable, even Nobel-Prize-level successes in various specific protein modeling tasks. 
However, those previous AI models still meet limitations, such as the difficulty in comprehending the grammar and semantics contained in protein sequences, and the inability to generalize across a wide range of protein modeling tasks. 
Recently, Large Language Models (LLMs) have emerged as a milestone in AI advances due to their remarkable language processing capability and unprecedented generalization capability. 
They are capable of promoting comprehensive progress in fields, rather than merely solving individual tasks. 
As a result, researchers have actively introduced powerful LLM techniques in promoting computational protein science, developing protein Language Models (pLMs) that skillfully grasp the foundational knowledge of proteins and can be effectively generalized to solve a diversity of sequence-structure-function reasoning problems.
While witnessing prosperous developments, it's necessary to present a systematic overview of computational protein science empowered by LLM techniques. 
First, we summarize existing pLMs into categories based on their mastered protein knowledge, i.e., underlying sequence patterns, explicit structural and functional information, and external scientific languages. 
Second, we introduce the utilization and adaptation of pLMs, highlighting their remarkable achievements in promoting protein structure prediction, protein function prediction, and protein design studies. 
Then, we describe the practical application of pLMs in antibody design, enzyme design, and drug discovery. 
Finally, we specifically discuss the promising future directions in this fast-growing field.

\end{abstract}

}

\maketitle

\IEEEdisplaynontitleabstractindextext

\IEEEpeerreviewmaketitle

\section{Introduction}

\begin{figure}[!t]%
  \centering
  \includegraphics[width=\linewidth]{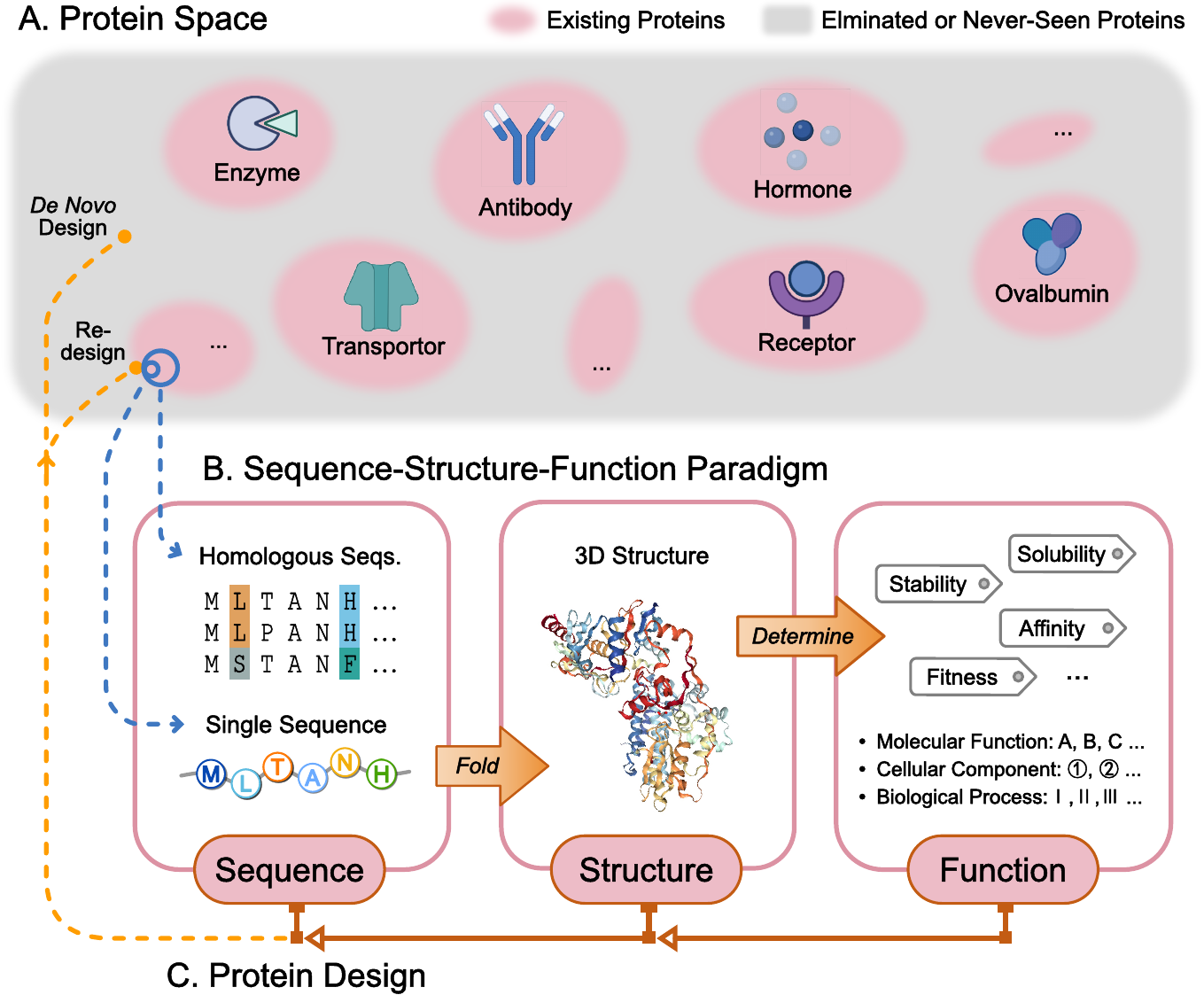}
  \captionsetup{font=small}
  \caption{Illustration of the Evolution and Sequence-Structure-Function Relationships. (\textbf{A}) The arrangement of amino acids forms a vast space of possible protein sequences. However, only a few proteins can survive through millions of years of evolution. (\textbf{B}) Valid amino acid sequences would fold into stable 3D structures and carry out proper functions. (\textbf{C}) Information flow within the sequence-structure-function paradigm can be leveraged in reverse, leading to the optimization of existing proteins or \textit{de novo} protein design oriented by desired functions.
  }
  \label{fig0}
\end{figure}

As the most foundational building blocks of life,  proteins play essential roles in almost all biological cellular processes~\cite{ferruz2022controllable,bepler2021learning}, such as metabolism, signal transduction, immune responses, etc. 
After long research, as illustrated in Figure \ref{fig0}, people have reached a limited understanding of the nature of proteins: 
Primarily, proteins adhere to the \textbf{sequence-structure-function} paradigm~\cite{anfinsen1973principles,redfern2008exploring} --- the \textit{amino acid (\textbf{AA})\footnote{In this work, the terms "amino acid" and "residue" are used interchangeably.} sequence of a protein indicates its three-dimensional structure, which in turn determines its function}. 
Moreover, proteins are shaped by the forces of \textbf{evolution} --- natural selection reserves protein sequences capable of folding into stable structures and fulfilling proper functions, while eliminating those that do not~\cite{maynard1970natural}.
Therefore, the protein sequence is widely acknowledged as the protein language~\cite{ofer2021language,bepler2021learning}, where the underlying arrangement patterns of AAs resemble "grammar" and the encoded structural and functional information mirrors "semantics." 
In this context, in the continued progression of scientific exploration, there are greater challenges in deciphering the protein language and applying rules of information flow among the protein sequence-structure-function. Computational protein science has emerged as a vitally important research field. 

With the ability to extract patterns and fit mappings from data, Artificial Intelligence (AI) techniques have been widely adopted in computational protein science, which has even driven groundbreaking results at the Nobel Prize level, 
with David Baker awarded for "\emph{computational protein design}" and Demis Hassabis \& John Jumper jointly awarded for "\emph{protein structure prediction}". 
Technically, the "AI for Protein" studies propose a variety of networks to accomplish different protein modeling tasks. 
For instance, UniRep~\cite{alley2019unified} trains a mLSTM model on unlabeled AA sequences to distill the fundamental features of a protein into a statistical representation. 
AlphaFold2~\cite{jumper2021highly} and RoseTTAFold~\cite{baek2021accurate} achieve a breakthrough in the accurate prediction of protein structures by exploiting the evolutionary information within multiple homologous sequences. 
DeepGOPuls~\cite{kulmanov2020deepgoplus} combines heuristic sequence similarity and a CNN model to predict protein functional annotations.
Variational Autoencoder (VAE) is employed for protein sequence generation~\cite{schmitt2022prediction} and backbone structure generation~\cite{eguchi2022ig}, which are important links in computational protein design. 
Although these AI-empowered protein modeling methods excel in their own specific tasks, they still have certain limitations. 
To be specific, traditional AI4Protein models, even some of their protein representation learning methods, cannot sufficiently understand the critical "grammar" and deep "semantics" within protein language. 
This limitation arises from their inferior capabilities in sequence processing and the integration of world knowledge.
Meanwhile, most existing protein modeling methods are designed for specific tasks and lack the generalization capability needed for multiple and even unseen tasks in the training stage.

Recently, Large Language Models (LLMs) have represented the forefront and peak of Artificial Intelligence, 
characterized by millions to billions of parameters and large-scale training based on extensive datasets. 
As is widely recognized, LLMs like BERT~\cite{devlin2018bert}, T5~\cite{raffel2020exploring}, GPT series~\cite{radford2018improving,radford2019language,brown2020language} and LLaMA~\cite{touvron2023llama} have achieved remarkable success and boosted the comprehensive advancements in various research fields, such as Natural Language Processing (NLP)~\cite{zhao2023survey}, Recommender Systems~\cite{fan2023recommender,qu2024tokenrec}, Healthcare~\cite{he2023survey}, and more. 
Essentially, LLMs spark with two unprecedented superiorities. 
\textit{First}, through large-scale pre-training from the extensive open-world knowledge, LLMs acquire foundational emergent capabilities, particularly in language understanding and generation. 
For text, a typical kind of sequence data, LLMs excellently grasp the arrangement patterns of tokens that conform to the basic grammar and express high-level semantics. 
\textit{Second}, LLMs exhibit unprecedented generalization capability with the support of techniques like probing, fine-tuning, prompting, etc. 
They are able to address a wide array of problems, moving beyond merely improving the prediction performance on individual specific tasks. 
Thus, building on the successful experience in natural language processing, LLMs have shown great potential for deciphering other "language sequences." 
Specialized LLMs have been developed to process genomic sequences~\cite{consens2023transformers}, chemical molecules~\cite{li2024empowering}, and especially --- the proteins. 

By combining powerful LLM techniques with abundant protein data, various \textbf{protein Language Models (pLMs)} are proposed to sufficiently grasp the foundational protein knowledge. 
More specifically, sequence-only pLMs (e.g, ESM-2~\cite{lin2023evolutionary}, ProtGPT2~\cite{ferruz2022protgpt2}, xTrimoPGLM~\cite{chen2024xtrimopglm}, ESM-MSA-1b~\cite{rao2021msa}) capture the valid AA arrangement patterns that have emerged over the course of evolution. 
pLMs that incorporate explicit structure and function information (e.g., SaProt~\cite{su2023saprot}, ESM-3~\cite{hayes2025simulating}) are effectively enhanced in protein understanding and generation. 
pLMs that learn natural language as well (e.g., ProLLaMA~\cite{lv2024prollama}, BioT5~\cite{pei2023biot5}) can understand a broad biomedical background and have the ability to follow textual instructions. 
Moreover, pLMs have been excellently generalized to the protein structure prediction, protein function prediction, and protein design problems --- those studies of sequence-structure-function reasoning have ushered inspiring progress simultaneously. 
For example, the representations learned by pLMs have empowered the accurate and fast inference from a single protein sequence to a 3D structure at atomic resolution~\cite{lin2023evolutionary}; 
the "pLM encoder - LLM decoder" framework unifies multi-task protein function prediction in a question-answering form~\cite{guo2023proteinchat}; 
the autoencoding and autoregressive pLMs well assist in protein redesign~\cite{notin2024proteinnpt} and \textit{de novo} protein design~\cite{madani2023large}, respectively. 
Taking all of those booming research trends into account, it's imperative to conduct a thorough review of pLMs and the latest advances in computational protein science promoted by them.

In this survey, we provide a systematic overview of computational protein science empowered by LLM techniques, so as to help researchers with an AI or biology background quickly understand relevant developments and gain insights. 
In particular, the rest of this paper is organized as follows: Section 2 presents a background of protein data profiles, AI for protein, and large language models. Section 3 categorizes the existing pLMs into sequence-based ones, structure-\&-function enhanced ones, and multimodal ones. Section 4 summarizes the utilization and adaptations of pLMs by considering the pending problems of protein structure prediction, protein function prediction, and protein design. Section 5 introduces some biomedical applications, including antibody design, enzyme design, and drug discovery. Section 6 provides a discussion on the current challenges and potential future directions. %
Our main contributions can be summarized as follows: 
\begin{itemize}[leftmargin=8pt, itemsep=0pt, topsep=1pt, parsep=1pt]
\item We conduct a comprehensive literature review on computational protein science in the era of LLMs, covering the foundational pLMs, the utilization and adaptation of pLMs, and some biomedical applications.
\item We present a systematic categorization for pLMs, emphasizing their learned knowledge of protein sequence, protein structure \& function, and external languages, and outlining the mainly employed approaches in building pLMs. 
\item We delve into the utilization and adaptation of pLMs, highlighting their impacts on the protein sequence-structure-function reasoning problems, and introducing the typical technical strategies. 
\item We describe some of the latest applications of pLMs, and discuss the prospective future directions in this field. 
\end{itemize}

To date, there are a limited number of existing surveys related to this topic. 
Regarding foundation models, Zhang et al.~\cite{zhang2024scientific} enumerate existing pLMs and categorize them by network architectures, and Hu et al.~\cite{hu2022protein} introduces the connection and progression between pLMs and structure prediction. 
Focus more on specific protein modeling problems, Unsal et al.~\cite{unsal2022learning} implement a benchmark for functional property learning of proteins based on language models, while Ferruz et al.~\cite{ferruz2022controllable} and Rufflo et al.~\cite{ruffolo2024designing} both report the controllable protein design with language models. 
In summary, they have not comprehensively encompassed the contents of foundation pLMs, downstream generalizations, and real-world applications, and their categorizations can be further improved by balancing the challenges in scientific explorations and technical strategies. 
Therefore, a comprehensive survey with systematic categorizations in this rapidly developing field is still in demand.

\section{Background}

In this section, we provide a brief review of the relevant background, including the biological basis and data profiles of proteins, the AI for protein studies, and large language models (LLMs).

\subsection{Biological Basis and Data Profiles}

\begin{figure*}[!t]%
  \centering
  \includegraphics[width=0.85\linewidth]{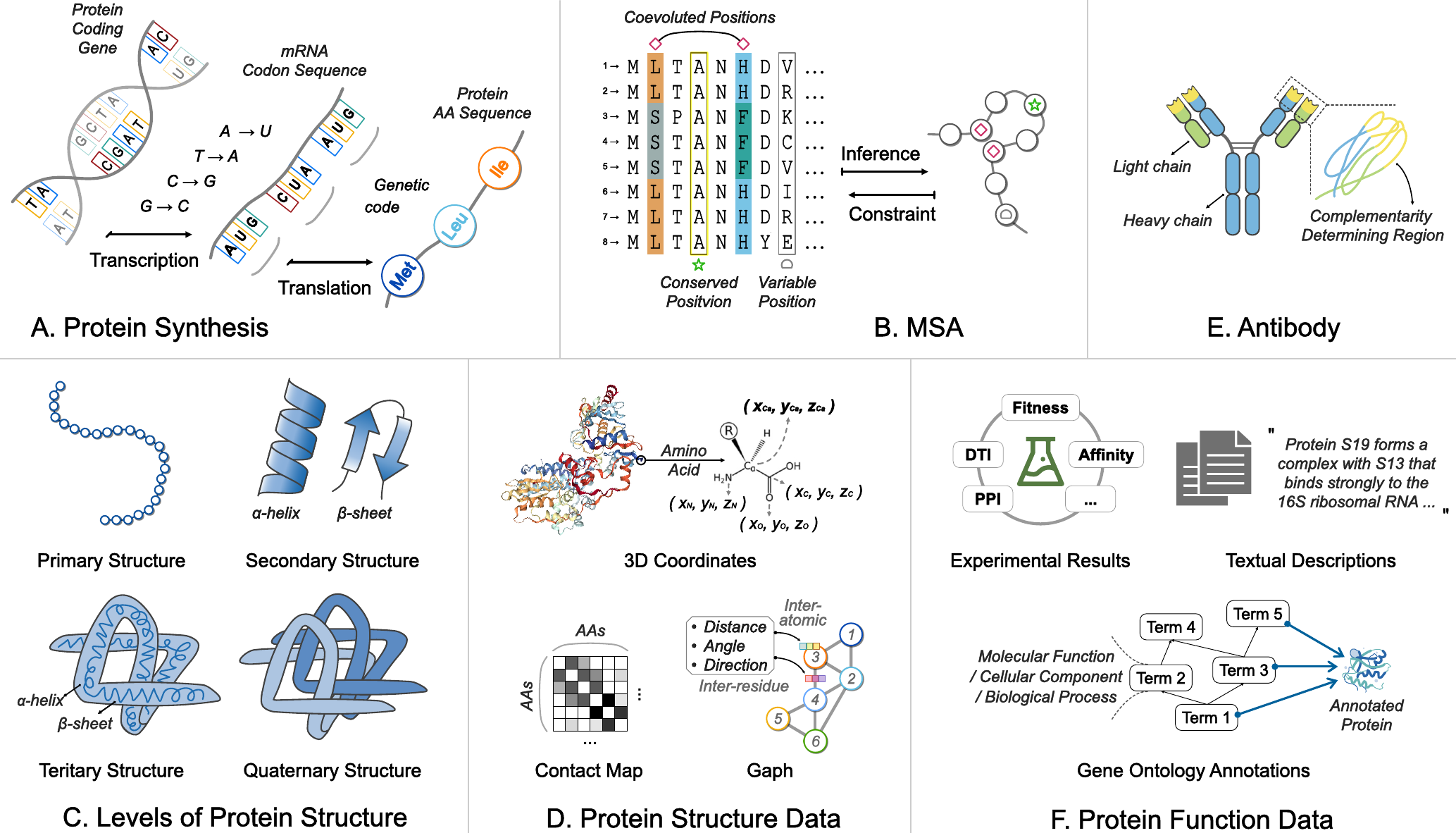}
  \captionsetup{font=small}
  \caption{Biological basis and data profiles. 
  (\textbf{A}) Protein synthesis mainly involves the transcription of protein-coding genes to mRNAs and the translation of codon sequences to AA sequences. 
  (\textbf{B}) Multiple Sequence Alignment (MSA) contains the evolutionary prior knowledge of proteins. Conserved positions are interpreted as core AAs for protein structure, as no changes have been allowed throughout the evolutionary process. Pairs of coevoluted positions indicate the spatial contacts of AAs, since mutations would occur and act synergistically to preserve the structure stable and unchanged. (\textbf{C}) Protein structure exhibits hierarchical organization. 
  (\textbf{D}) Protein structure can be described in several forms. 3D coordinates of atoms trustfully record the experimentally determined protein conformation. A 2D distance map conveys the proximity between all possible AA pairs. Furthermore, we can build specific graphs to describe detailed structural characteristics, where the interatomic or inter-residue distances, angles, and directions are encoded as node and edge features. 
  (\textbf{E}) An antibody is a \textsf{Y}-shaped protein composed of two heavy and two light chains. At the top of the "\textsf{Y}"'s arms, complementarity-determining regions (CDRs) are polypeptide segments that make up the antigen binding site. 
  (\textbf{F}) Protein function is described in multiple formats, such as lab-generated labels, Gene Ontology annotations, and textual documents.}
  \label{fig2}
\end{figure*}

Understanding the synthesis, evolution, structure, and function of proteins would pave the way for advances in biology and medicine.
Over the years, increasing attention has been paid to studying proteins for various tasks in science. 
For example, many scientists from the wet lab have conducted comprehensive wet experiments through advanced biotechnology (e.g., mass spectrometry~\cite{de2007mass}, X-ray crystallography~\cite{read2011new}, electron microscopy~\cite{henderson2012outcome}, and deep mutational scanning~\cite{fowler2014deep}), significantly contributing to the accumulation of a substantial volume of high-quality data~\cite{suzek2015uniref,wwpdb2019protein,ashburner2000gene}. 
Thus, we start by providing some background biological knowledge and representative data formats in Figure \ref{fig2}.

In the wild, protein synthesis involves two processes: transcription and translation. 
Genetic information is transcribed from protein-coding genes into messenger RNAs (mRNA), which are subsequently translated into proteins~\cite{crick1970central}. 
Specifically, the mRNA sequence is composed of four types of nucleotides, i.e., adenine (\texttt{A}), uracil (\texttt{U}), cytosine (\texttt{C}), and guanine (\texttt{G}), 
while a codon refers to a subsequence of three consecutive nucleotides, such as "\texttt{AUG}", "\texttt{AAG}", etc.
According to the genetic code~\cite{koonin2009origin}, the 20 standard AAs are encoded by 61 codons. 
For example, Methionine (abbr. as \texttt{Met} or \texttt{M}) is specified by the codon "\texttt{AUG}", while Lysine (abbr. as \texttt{Lys} or \texttt{K}) can be translated from the codons "\texttt{AAA}" or "\texttt{AAG}". Through the sequential translation of codons and post-translational modifications, a protein is primarily constituted by a sequence of amino acids, which is also called the polypeptide chain. We can denote it as $ X = \left[ x_1, x_2, ..., x_L \right] $, where $ L $ refers the sequence length, and $ x_i \left( 1 \leq i \leq L \right) $ represents each individual amino acid.

The total number of possible amino-acid sequences can easily exceed orders of billions, yet only a minute fraction has existed on earth since the origin of life~\cite{maynard1970natural}. Sequences that fail to fold into stable structures and exhibit effective functionality are more likely to be eliminated in the course of evolution, while the survived natural protein sequences would reflect evolutionary favored patterns. To be specific, evolution gives rise to the emergence of protein families that share significant structural and functional similarities~\cite{chothia1992one}: Random genetic mutations occur in an ancestral protein over time, and certain descendant variants survive through the process of natural selection; Beneficial mutations accumulate in the ancestral protein within the population across generations, which leads to the adaptation and divergence of the protein family. In computational analysis, it's a common practice to retrieve homologous sequences and create multiple sequence alignment (MSA) to incorporate the prior knowledge of protein evolution. Concretely, an MSA comprising $M$ sequences of length $L$ can be encoded as a matrix $X \in \mathbb{R}^{M \times L}$, where each entry $x_{m, l} \left( 1 \leq m \leq M, 1 \leq l \leq L \right)$ represents the amino acid identity of sequence $m$ at position $l$. By comparisons, we can identify the conserved positions that remain unchanged, the variable positions that allow different kinds of mutations, and the co-evoluted positions that are changed synchronously as constrained by structural contacts.

After the synthesis of polypeptide chains, protein folding is a fundamental physical process that transforms the unstable sequential conformations into more ordered three-dimensional structures~\cite{dill2012protein}. The primary structure of a protein refers to the linear arrangement of AAs in the polypeptide chain, which holds great importance in encoding the protein's 3D structure and biological function from the very beginning. Then, the secondary structure refers to highly regular local sub-structures along the polypeptide chain, typically including $\alpha$-helices and $\beta$-sheets. The tertiary structure refers to the 3D structure formed based on a single polypeptide chain, where a series of $\alpha$-helices and $\beta$-sheets are folded compactly through non-specific hydrophobic interactions. The quaternary structure describes the 3D structure formed by aggregating at least two individual polypeptide chains, which function together as multimers. Protein Data Bank (PDB)~\cite{wwpdb2019protein} is an authoritative database that collects experimentally determined protein structures, while "\textit{pdb}" and "\textit{mmcif}" are widely acknowledged file formats that provide detailed descriptions of the 3D structures of proteins. A structure file comprehensively records the \textit{(x, y, z)} coordinates for each atom constituting all amino acids. Then, the structure of a protein can be redescribed by executing various data processing schemes. For example, we can calculate the relative distance, angle, and direction among atoms or AAs, and build a graph that represents AAs as nodes and their spatial relations as edges~\cite{jing2020learning,gao2022pifold}.

A fundamental principle in molecular biology asserts that the structure of a molecule inherently determines its function~\cite{redfern2008exploring}. For example, antibodies are the \textsf{Y}-shaped proteins employed by the immune system to specifically identify and neutralize pathogenic organisms, like bacteria and viruses. The two arms in the "\textsf{Y}"-like structure are crucial as hosting antigen-binding sites that recognize and bind to antigens. Then, the detailed conformation of antigen-binding sites further determines the antibody's functional affinity for clearing specific pathogens. 
Beyond the antibodies, there are broader facts that demonstrate the complexity of protein function mechanisms. Structural differences among proteins correlate strongly with functional diversity, and even homologous proteins with similar structures can exhibit significant variations in the fitness landscape~\cite{romero2009exploring}. Considering the great complexity, protein function is presented by a wide range of data formats, and we summarize them as
experimental results, manual annotations, and textual descriptions: 

\begin{itemize}[leftmargin=8pt, itemsep=0pt, topsep=1pt, parsep=1pt]
\item First, scientists conduct laboratory experiments to measure functional properties (e.g., fitness, stability), identify functional sites (e.g., signal peptides, B-cell epitope), and detect biomolecular interactions (e.g., drug-target interaction, protein-protein interaction). All kinds of results are collected as labels in datasets. 
\item Second, proteins can be annotated with authoritative classes. Gene Ontology (GO) knowledgebase~\cite{ashburner2000gene} organizes a hierarchical framework of terms that describe various classes of molecular functions, cellular components, and biological processes. Each protein is manually tagged with multiple GO terms, facilitating a granular and systematic exploration of protein functions. Besides, Enzyme Commission (EC) number is an authoritative taxonomy that assigns a unique code to each enzyme based on their catalyzed chemical reactions. Like "EC 2.7.6.1", an enzyme code is composed of a beginning string "EC" and four numbers separated by dots, denoting a progressively finer classification of the enzyme. 
\item Third, academic publications serve as an essential resource, providing comprehensive empirical evidence and theoretical analyses for understanding protein functions. 
\end{itemize}

\subsection{AI for Protein Science}

There are essential protein modeling problems centered in the sequence-structure-function paradigm, including \emph{protein representation learning, structure prediction, function prediction}, and \emph{protein design}. In recent years, artificial intelligence has significantly advanced these studies. 

As one of the most representative techniques in protein modeling, protein representation learning aims to extract latent knowledge of proteins from extensive data and encode individual proteins into vector representations. 
Then, the learned protein knowledge can be used for various downstream applications. As summarized by Wu et al.~\cite{wu2022survey}, existing protein representation learning methods can be categorized into three groups: sequence-based protein encoders, structure-based protein encoders, and sequence-structure co-modeling methods. 
First, UniRep~\cite{alley2019unified}, CARP~\cite{yang2024convolutions}, and ProtTrans~\cite{elnaggar2021prottrans} train LSTM, CNN, and Transformer models on single AA sequences, while ESM-MSA-1b~\cite{rao2021msa} propose customized axial Transformers for multiple sequence alignments (MSAs). 
These encoders are mainly pre-trained using the mask language modeling objective. 
Second, GearNet~\cite{zhang2022protein} and GVP-GNN~\cite{jing2020learning} utilize message-passing graph neural networks (GNNs) or geometric-aware GNNs on protein structure graphs. These encoders learn by contrasting sampled structures and incorporating specific supervision on structural characteristics. 
Third, LM-GVP~\cite{wang2022lm} and ESM-GearNet~\cite{zhang2023systematic} introduce novel strategies that combine sequence models and GNNs by modifying the network architectures and learning objectives. 
It is worth noting that protein language models (pLMs) are closely connected with protein representation learning. Certain pLMs, particularly encoder-only models, can be acknowledged as protein representation methods. In addition, pLMs are commonly integrated within protein representation methods as a crucial component, working alongside modules that extract structural or functional features from proteins.

Protein structure prediction is designed to infer the 3D structure of a protein according to its amino acid sequence. At a lower resolution, protein structure prediction is defined as secondary structure prediction or contact map prediction: An AA sequence is mapped to a series of secondary structure elements (i.e., $\alpha$-helix, $\beta$-sheet, or other conformation); Each pair of residues in the sequence is mapped to whether they are in contact. 
Furthermore, AI models have made substantial progress in predicting protein structures at the atomic resolution, i.e., inferring accurate 3D coordinates. 
Prominent methods, such as AlphaFold2~\cite{jumper2021highly}, RoseTTAFold~\cite{baek2021accurate}, ColabFold~\cite{mirdita2022colabfold}, etc., leverage evolutionary information to achieve unprecedented performance. 
These methods typically involve a two-step process: first, they search for homologous multiple sequence alignments (MSAs) and structure templates according to the input sequence; then, they generate coordinates for all atoms through well-designed networks that collectively analyze relationships within and between the 1D sequence, 2D contact, and 3D conformation. 
Despite achieving near experimental accuracy, these methods are reliant on MSAs and may use known structures from homologous proteins as references, which deviates slightly from the goal of capturing the underlying rules that govern polypeptide folding. 
In order to overcome this deficiency, efforts have been dedicated to single-sequence protein structure prediction. Certain pLM-based methods, such as ESMFold~\cite{lin2023evolutionary}, trRosettaX-Single~\cite{wang2022single}, and HelixFold-Single~\cite{fang2023method}, demonstrate competitive performance comparable to AlphaFold2 and RoseTTAFold while free of the prior co-evolution information.

Protein function prediction aims to conclude the functional knowledge of proteins based on their AA sequences or 3D structures. Given the diverse nature of protein functions, protein function prediction encompasses various concrete tasks. First, when considering each input protein as a whole, AI techniques are employed to predict experimentally measured properties (e.g., stability, fluorescence, fitness)~\cite{dallago2021flip} or manually curated annotations (e.g., gene ontology annotations, enzyme classification numbers)~\cite{kulmanov2020deepgoplus,fernandez2023exploring}.
Second, proteins perform specific functions through the finer functional sites, such as signal peptides, complementarity-determining regions, ligand-binding domains, etc. Consequently, AI models are developed to identify whether each residue belongs to a specified functional region~\cite{liu2024protein,yuan2024genome}. Third, proteins generally don't function in isolation but rather interact with biomolecules (e.g., DNA, RNA, drug, another protein) to carry out their functions. AI models are designed to determine whether interactions occur or to reveal interaction details based on pairwise protein-molecule inputs~\cite{singh2023contrastive,yu2023unikp}. 
In addition, a cutting-edge research spot involves enabling multi-task learning within a unified framework, which eliminates the need for numerous individual models. Efforts are made to develop ChatGPT-like systems that allow users to upload proteins and engage in question-answer conversations to gain insights~\cite{guo2023proteinchat,wang2024protchatgpt,abdine2024prot2text}.

Protein design aims to create new proteins with functions surpassing existing proteins by discovering optimized or novel sequences~\cite{notin2024machine}. It is required to precisely identify specific regions within the vast protein sequence space that give rise to desired functions. Depending on whether starting from known proteins or scratch, relevant studies can be classified into two groups: protein redesign and \textit{de novo} protein design. During protein redesign, mutations are introduced to existing proteins to enhance their functional properties. Therefore, limited protein space surrounding existing proteins is explored. Autoenconding pLMs trained by masked language modeling excel at predicting highly probable mutations favorable in evolution, inspiring "new family members" that are directionally evolved~\cite{meier2021language,notin2024proteinnpt}. Dissimilarly, \textit{de novo} protein design involves autonomously discovering optimal amino acid sequences that satisfy a given design objective, enabling the exploration of regions not previously seen in evolutionary history within the extensive protein space. Most \textit{de novo} protein design methods divide the workflow into two stages~\cite{kortemme2024novo}: 1) Structure generation models (e.g., RFDiffusion~\cite{watson2023novo} and Chroma~\cite{ingraham2023illuminating}) create the protein's structural backbone without a defined sequence; 
2) Inverse folding models (e.g., ESM-IF~\cite{hsu2022learning} and ProteinMPNN~\cite{dauparas2022robust}) recover effective sequences based on the backbone atom coordinates. 
Besides, certain studies consider \textit{de novo} protein design as a single-step computational task, i.e., protein sequence generation. Autoregressive pLMs (e.g., ProGen~\cite{madani2023large} and ProtGPT2~\cite{ferruz2022protgpt2}) and diffusion frameworks (e.g., EvoDiff~\cite{alamdari2023protein}) can generate protein sequences that exhibit similar physicochemical properties to natural proteins but have low homology to existing proteins.

\subsection{Large Language Models (LLMs)}
Recently, the development of LLMs has unleashed a great surge in AI~\cite{brown2020language,zhao2023survey}. Typically, LLMs with million or billion-level parameters are pre-trained on extensive data, demonstrating surprising abilities in various NLP tasks like language understanding, text generation, etc. Due to their powerful understanding and generation capability, LLMs have been widely adopted in various fields, including data mining~\cite{fan2023recommender,fan2024graph}, healthcare~\cite{he2023survey} and molecule science~\cite{li2024empowering,liu2024moleculargpt}, by fine-tuning or prompting with domain-specific datasets. According to the architecture and functionality, LLMs are categorized into three primary types: \emph{Autoencoding models}, \emph{Autoregressive models} and \emph{Sequence-to-Sequence models}.

As a typically Autoencoding model, BERT~\cite{devlin2018bert}
transforms input text into a high-dimensional latent space to capture the contextual semantic information of the text. One of the key features of Autoencoding models is their bi-directionality nature, which allows them to process both the preceding and following context. This feature enables Autoencoding models to better understand and represent the input data, which is important in tasks like sentiment classification and language translation.
Different from Autoencoding, Autoregressive models such as GPT~\cite{radford2018improving}
family processes the input text in a left-to-right manner. They generate the next token by predicting based on the context of previous tokens. This autoregressive generation fashion allows them to generate coherent and contextually relevant text in tasks such as creative writing and code generation. In addition, Sequence-to-Sequence models like T5~\cite{raffel2020exploring} take conditional generation as the core idea. Specifically, they use an encoder to capture the meaning of the input text and a decoder to generate the output based on encoding information. This architecture makes Sequence-to-Sequence models a unified framework to handle various NLP tasks flexibly. 
It is worth noting that the performance of LLMs follows a scaling law~\cite{kaplan2020scaling}.
Specifically, the performance of LLMs has been observed to scale predictably with increases in model size and the amount of training data~\cite{brown2020language}. 

One typical method for adopting LLMs to various downstream predictions is fine-tuning the LLMs on specific datasets. That is to say, the pre-trained parameters are subsequently trained to learn more specific tasks. However, full model fine-tuning still requires considerable data samples, costs extensive computing resources, and takes quite some time. To overcome these limitations, recent studies explore Parameter-Efficient Fine-Tuning (PEFT) to adopt partial parameters instead of fine-tuning the LLMs extensively~\cite{hu2021lora,dettmers2023qlora}. For example, Low-Rank Adaptation (LoRA)~\cite{hu2021lora} allows for the effective adaptation of LLM by only updating a few low-rank matrices, thus substantially decreasing the number of parameters while preserving high-performance levels.

To further reduce the reliance on training data, prompt learning recently emerged as a popular paradigm for adopting LLMs to various tasks, with the input carefully designed to guide LLMs in generating the desired output without intensive parameter updates. In-context learning is a commonly used prompt learning method that provides a few task demonstrations within the prompt to instruct the LLMs to perform downstream tasks. Furthermore, Chain-of-Thought (CoT)~\cite{wei2022chain} is another specific prompting technique. The key idea of CoT prompting is to annotate intermediate reasoning steps into the prompt to enhance the reasoning ability of LLMs.
While these prompting methods achieve great success, such manually designed prompts typically face discrete optimization challenges, such as laborious trial and error in finding suitable prompts. To solve this challenge, soft prompt tuning~\cite{lester2021power} is introduced, where prompts are not fixed text but rather continuous, trainable embeddings, allowing a more nuanced and flexible prompt design. 

Furthermore, recent studies have expanded the utility of LLMs by integrating them with other modalities or external knowledge bases to enhance their performance and versatility. For example, models like CLIP~\cite{radford2021learning} utilize contrastive learning to comprehend images through textual descriptions, showcasing exceptional abilities in cross-modal understanding. PaLM-E architecture~\cite{driess2023palm} integrates the specific encoder and the uni-decoder hierarchically with the help of an internal projector~\cite{pei2024leveraging}, enabling it to generate contents based on the understanding of visual and language inputs. Similarly, BLIP-2~\cite{li2023blip} introduces a frozen image encoder with a pre-trained text decoder, where a lightweight Querying Transformer (Q-Former) module is developed to bridge the modality gap. 
Furthermore, without updating the LLM backbone, Retrieval-Augmented Generation (RAG) techniques have emerged as an effective approach to enhance the understanding and generation capabilities of LLMs by retrieving external knowledge database~\cite{fan2024survey}.

\section{Pre-trained Protein Language Models}

In recent years, LLM techniques have been employed not only in processing natural languages but also in deciphering various domain-specific "languages". Substantial advancements have been made in protein language models (\textbf{pLMs}). In this section, we systematically review existing pLMs, categorizing them as sequence-based, structure-and-function-enhanced, and multimodal models.

\subsection{Sequence-based pLMs}

General LLMs capture the interdependencies among subword tokens and acquire a profound understanding of the grammar and semantics of text. Similarly, sequence-based pLMs capture the mutual dependencies among amino acid (AA) tokens, extract the favorable sequence patterns, and grasp implicit structural and functional information. Sequence-based pLMs can be further distinguished as single-sequence-based ones and multiple-sequence-based ones. The former describes each protein by the corresponding AA sequence, while the latter possesses an idea of retrieval augmentation, describing each protein with multiple related sequences in evolution or synthesis. Table \ref{tab1} presents a comprehensive summary, outlining the input data, network architecture, and pre-training objective of each pLM.

\subsubsection{Single-Sequence-based pLMs}

By taking each amino acid sequence as a sentence, the valuable practice of building general LLMs has been extended to the development of pLMs. We have witnessed the emergence of autoencoding pLMs in the BERT-style, autoregressive pLMs following the GPT approach, and sequence-to-sequence pLMs resembling T5 or GLM. 

\begin{figure*}[!t]
  \centering
  \includegraphics[width=0.9\textwidth]{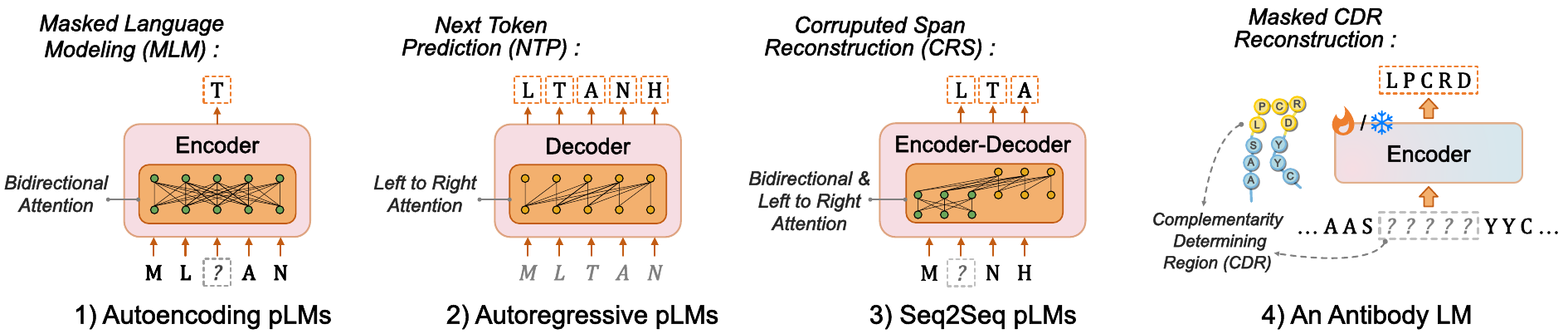}
  \captionsetup{font=small}
  \caption{Typical \textbf{single-sequence-based pLMs}. \textbf{1-3)} When considering individual amino acid sequences as "sentences", pLMs follow the general approaches of autoencoding, autoregressive, and sequence-to-sequence as well. \textbf{4)} Masked CDR reconstruction is a novel pre-training objective that incorporates the inherent characteristics of antibodies into mask language modeling. }
  \label{fig3}
\end{figure*}

As shown in Figure \ref{fig3}-1, \textbf{Autoencoding pLMs} employ the Transformer encoder with bidirectional attention and undergo pre-training via the masked language modeling (MLM) objective. They are skilled in encoding the context of protein sequences into informative representations. ESM-1b~\cite{rives2021biological} is a pioneering study that executes self-supervised learning on 250 million protein sequences spanning evolutionary diversity. Technically, ESM-1b employs the architecture and pre-training procedure of RoBERTa~\cite{liu2019roberta} almost as-is, thereby driving the model to extract the residue–residue dependencies latent in the extensive sequence data. 
Notably, it is revealed that the structural residue-residue contacts can be inferred from the pLM-produced protein sequence representations by linear projections. The significant correlation between the actual distance map and the extracted inter-residue dependencies underscores a valuable consistency between the nature of proteins and the framework of language models. 
With a similar idea, ProtTrans~\cite{elnaggar2021prottrans} involves several other representative autoencoding LMs
to learn the protein language, resulting in ProtBERT, ProtAlbert, and ProtElectra.
In transfer learning evaluations, those single-sequence-based presentations perform competitively with classic methods that take in multiple sequence alignments. 
This finding implies that pLMs have captured some of the grammar of the protein language, even without direct exposure to evolutionary information.

Subsequently, to explore the boundaries of sequence-based protein representation learning for optimal performance, the next-generation study of ESM-1b, i.e., ESM-2~\cite{lin2023evolutionary}, implements the scaling law of LLMs. With the scale of parameters increasing from 8 million to 15 billion, significant improvements are observed in the fidelity of protein sequence modeling, leading to the emergence of protein structure knowledge at the atomic level within the learned representations. As of December 2024, the ESM team has made the latest generation, ESM Cambrian (ESM C)~\cite{esmc}, available. It is claimed that ESM C establishes a new state-of-the-art performance for protein sequence modeling, achieving linear scaling across orders of magnitude in parameters. With improved training efficiency, each ESM C model could match or even greatly exceed the performance of previous larger ESM-2 models. 

Besides, efforts have been made to improve autoencoding pLMs from different points of view. To better capture co-evolutionary information in inter-residue co-variation, He et al.~\cite{he2021pre} introduce a novel pre-training objective called pairwise masked language modeling (PMLM). For extrapolation of longer proteins and protein complexes, LC-PLM~\cite{wang2024long} is developed using a novel compute-efficient network architecture called bidirectional Mamba with shared projection layers (BiMamba-S)~\cite{qu2024survey,qu2024ssd4rec}. To mitigate computational resource requirements, ProtFlash~\cite{wang2023deciphering} employs a mixed chunk attention strategy with linear complexity, while DistillBERT~\cite{geffen2022distilprotbert} is proposed as a distilled variant of ProtBERT. 
To find a balance between performance and efficiency in the trend of scaling up pre-training, AIDO.Protein~\cite{sun2024mixture} explores the Mixture-of-Experts (MoE) model in computational protein science for the first time. It serves a critical link in the AI-Driven Digital Organism (AIDO)~\cite{song2024toward} system, which aims to integrates multi-scale foundation models to predict, simulate, and program biology at all levels. 
Besides, DPLM~\cite{wang2024diffusion} is a versatile protein language model under a discrete diffusion framework~\cite{liu2023generative}. While maintaining the bi-directional receptive field, the key ingredient of autoencoding modeling, an additional diffusion pre-training process makes DPLM possess a strong and scalable generative power. 

As shown in Figure \ref{fig3}-2, \textbf{Autoregressive pLMs} leverage Transformer decoder to process the protein sequence left-to-right and conduct pre-training using the next token prediction (NTP) objective. They are good at generating protein sequences that adhere to the patterns favored by evolution
Building upon the approach of GPT-2, ProtGPT2~\cite{ferruz2022protgpt2} is capable of generating protein sequences that exhibit natural amino acid propensities yet relate distantly to naturally occurring sequences. 
Across the progression from GPT-2 to GPT-3 and subsequent iterations, general LLMs have consistently exhibited a trend of scaling up, which also applies to pLMs. 
RITA~\cite{hesslow2022rita} encompasses a suite of models ranging from 85 million to 1.2 billion parameters, which are trained on over 280 million protein sequences. ProGen2~\cite{nijkamp2023progen2} models are enlarged to 6.4 billion parameters and are trained on an extensive collection of over one billion protein sequences. Moreover, it is observed that even the largest ProGen2 model still exhibits underfitting, indicating the potential for further improvements in capturing the intrinsic distribution of natural protein sequences. 

As illustrated in Figure \ref{fig3}-3, \textbf{Sequence-to-sequence pLMs} employ either the encoder-decoder or non-causal decoder architecture and undergo pre-training with the corrupted spans reconstruction (CSR) objective. These models combine autoencoding and autoregressive traits, can encode the input sequence, and perform conditional generation accordingly. 
In ProtTrans~\cite{elnaggar2021prottrans}, ProtT5 is a collection of pLMs inspired by the original T5 series. While scaling up to 3 billion and 11 billion parameters, ProtT5 models produce highly informative protein representations that outperform the smaller autoencoding pLMs such as ProtBERT. 
Following ProtTrans, Ankh~\cite{elnaggar2023ankh} is another empirical study that explores data-efficient, cost-effective, and knowledge-guided optimization of pLMs. Using ProtT5 as the baseline, Ankh comprehensively investigates potential factors that could affect the performance of pLMs, covering more than twenty ablation experiments that compare detailed strategies in token span corruption, encoder-decoder architecture, and data sources. 
Besides, xTrimoPGLM~\cite{chen2024xtrimopglm} is also proposed to process the protein representation and protein generation uniformly. xTrimoPGLM leverages General Language Model (GLM)~\cite{du2021glm} as its backbone architecture and explores the joint optimization of MLM and CSR objectives, resulting in successful pLM pre-training at the scale of approximately 100 billion parameters and 1 trillion tokens as never before. 

In contrast to general proteins, antibodies (i.e., immunoglobulins) are specialized Y-shaped proteins with immune functions. As illustrated in Figure \ref{fig2}-E, antibodies are characterized by the two heavy chains and two light chains in their sequences. Then, complementarity-determining regions (CDRs) are vitally important components within antibody sequences that dictate the specific antigen-binding sites. 
Considering the critical medical significance and distinctive sequence organization of antibodies, \textbf{antibody language models} have been developed specifically. 
In early attempts, AntiBERTy~\cite{ruffolo2021deciphering}, AntiBERTa~\cite{leem2022deciphering}, and AbLang~\cite{olsen2022ablang} employed the classic autoencoding approach of BERT or RoBERTa to decipher antibody sequences. 
Furthermore, the pre-training objective of antibody LMs can be enhanced by incorporating relevant biological knowledge. As shown in Figure \ref{fig3}-4, AbBERT~\cite{gao2023pre} models are pre-trained by reconstructing masked CDR spans, thereby acquiring antibody-specific insights.
ReprogBert~\cite{melnyk2023reprogramming} investigates cross-language adaptation using limited data, reprograming an English BERT model into an antibody LM by training amino acid vocabulary embeddings with the masked CDR reconstruction objective. 
Besides, autoregressive and sequence-to-sequence antibody LMs have also been developed. For example, IgLM~\cite{shuai2023iglm} enables the generation of antibody sequences conditioned on prefix tag tokens that indicate the chain type and the origin species, paired-IgGen (p-IgGen)~\cite{turnbull2024p} is a generative LM pre-trained on both paired and unpaired heavy-light chain antibody sequences, and pAbT5~\cite{chu2023generative} exhibits the "translation" capability that generates light-to-heavy-chain or heavy-to-light-chain sequences. As learning the distinct language of antibodies, these models effectively contribute to the controllable design of antibodies.

\subsubsection{Multiple-Sequences-based pLMs}

Retrieval is a fundamental data mining technique designed to understand input queries and extract relevant information to assist in analysis~\cite{fan2024survey}. In computational protein science, retrieval has been a well-established concept for decades. 
Many significant analyses are conducted based on multiple biologically related sequences, aiming to leverage prior evolutionary knowledge, such as the co-variation between residues. Driven by the concept of retrieval and the practice in large-scale pre-training, multiple-sequences-based pLMs have been developed as well.

\begin{figure}[!t]%
  \centering
  \includegraphics[width=0.8\linewidth]{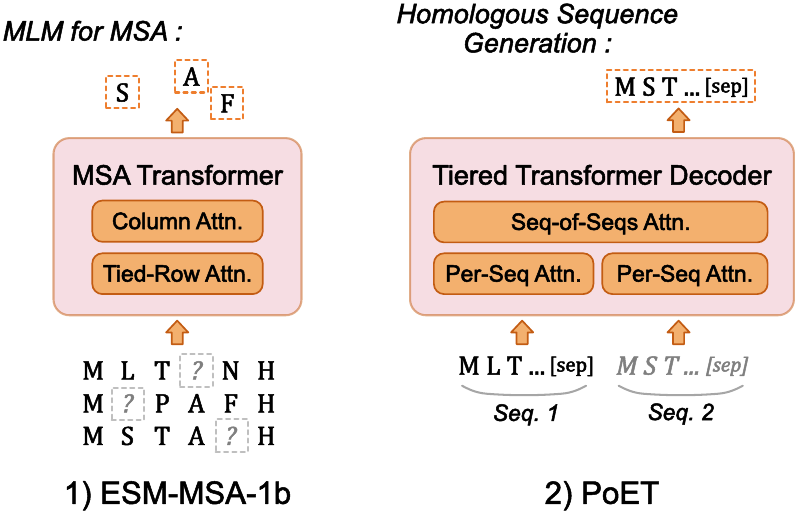}
  \captionsetup{font=small}
  \caption{Typical \textbf{multiple-sequences-based pLMs}. \textbf{1)} ESM-MSA-1b, a representative MSA-based pLM, incorporates bidirectional tied-row attention and column attention within each MSA Transformer block, thereby capturing co-evolution features within the 2D input. \textbf{2)} PoET is an autoregressive model specifically designed to learn the distribution over protein families. It accepts multiple sequences as input without the need for alignment and can generate sets of homologous proteins.}
  \label{fig4}
\end{figure}

Traditional searching tools like HHblits~\cite{remmert2012hhblits} and MMseqs2~\cite{steinegger2017mmseqs2} are widely employed to retrieve and align protein families from extensive databases. The resulting multiple sequence alignments (MSAs) can greatly contribute to investigating the evolutionary relationships between proteins. As shown in Figure \ref{fig2}-B, MSAs facilitate the identification of conserved, variable, or co-evolution regions within the protein sequences, offering valuable insights into their structural and functional implications. 
ESM-MSA-1b~\cite{rao2021msa} is the first pLM designed to operate on MSAs. As illustrated in Figure \ref{fig4}-1, ESM-MSA-1b incorporates bidirectional tied-row attention and column attention within each MSA Transformer block, thereby capturing the interdependencies among residues and across sequences. In evaluations, ESM-MSA-1b demonstrates great performance in unsupervised protein structure learning, surpassing single-sequence-based pLMs 
while with far fewer parameters. These findings meet the motivation that structural constraints can be effectively inferred from the patterns of protein sequences.

Moreover, as well-established protein structure prediction methods (e.g., AlphaFold2) rely heavily on MSAs and exhibit inadequacy when confronted with orphan proteins lacking adequate homologous sequences, there is an increasing demand to generate "pseudo" proteins based on existing MSAs for augmentation. MSA2Prot~\cite{ram2022few} is an MSA-to-protein Transformer that can generate individual protein sequence variants based on a learned encoding of input MSA. Then MSA-Augmenter~\cite{zhang2023enhancing-msa} leverages the tied-row \& column attention mechanism to generate an arbitrary number of co-evolutionary sequences. 
Besides, MSAGPT~\cite{chen2024msagpt} involves a flexible MSA decoding framework.
By employing a 2D positional encoding scheme that describes the complex evolutionary patterns, MSAGPT empowers the general 1D Transformer decoder for MSA processing.

However, from another point of view, the efficacy of MSA-based pLMs is usually hindered by the inadequacy of fundamental alignment algorithms, which can introduce errors such as long insertions and gappy regions. Even worse, the alignment errors tend to increase, and the computational efficiency decreases when handling longer sequences. 
To avoid these problems, certain pLMs are designed to operate on multiple sequences without the need for alignment. 
As shown in Figure \ref{fig4}-2, PoET~\cite{truong2024poet} contains a tired sequence-of-sequences causal attention mechanism to generate sets of homogeneous protein sequences. Besides, inspired by the recent advancement of Mamba~\cite{qu2024survey}, ProtMamba~\cite{sgarbossa2024protmamba} efficiently handles significantly long contexts, even encompassing hundreds of sequentially concatenated protein sequences.

In addition to retrieving homogeneous AA sequences, proteins can be supplemented by other biological sequences involved in the synthesis process. As introduced in Figure \ref{fig2}-1, each AA sequence naturally originates from an mRNA codon sequence, where genetic information can be exploited to build enhanced pLMs.
CaLM~\cite{outeiral2024codon} is designed to generate representations of codon sequences, providing valuable signals for protein engineering applications. 
cdsBERT~\cite{hallee2023cdsbert} introduces a pipeline 
to enhance the capability of pre-trained ProtBERT models by introducing codon awareness. 
Furthermore, post-translational modifications (PTMs) represent a covalent process that modifies proteins after their synthesis, playing a vital role in increasing proteins' structural and functional diversity. PTM-Mamba~\cite{peng2024ptm} stands as the first PTM-aware pLM, encompassing representations of both amino acids and PTM tokens.

\definecolor{navy}{RGB}{11, 11, 255}
\definecolor{pink}{RGB}{224, 33, 137}
\definecolor{gray}{RGB}{238, 238, 238}

\begin{table*}[!t]\scriptsize
\captionsetup{font=small}
    \caption{\textbf{Sequence-based pLMs}. We present the resource link, employed corpora, input data format, backbone architecture, scale of parameters, and pre-training objective for \textit{single-sequence-based pLMs} and \textit{multiple-sequences-based pLMs}. All abbreviations are explained in the footnote \textbf{*}, and important databases are also described in footnotes 1 to 8. In the context of "Architecture", LLM or pLM names devoid of color indicate training corresponding networks from scratch, names in \textcolor{navy}{blue} represent using pre-trained models while keeping parameters frozen, and names in \textcolor{pink}{pink} indicate using pre-trained models with parameters trainable.}
    \begin{threeparttable}[b]
    \centering
    \rowcolors{2}{gray}{white}
    \begin{tabular}{ccccccc}
    \toprule
    \textbf{Method} & \textbf{Corpora} & \textbf{Input} & \textbf{Architecture} & \textbf{\#Parameter} & \textbf{Objective} \\
    \midrule
    \href{https://huggingface.co/facebook/esm1b_t33_650M_UR50S}{ESM-1b}~\cite{rives2021biological} & \href{https://www.uniprot.org/uniref?query=*&facets=identity%3A0.5}{UniRef50}~\cite{suzek2015uniref}\textsuperscript{1} & AA Seq. & RoBERTa~\cite{liu2019roberta} & 650M & MLM \\
    \href{https://huggingface.co/facebook/esm2_t33_650M_UR50D}{ESM-2}~\cite{lin2023evolutionary} & UniRef50 &  AA Seq. & RoBERTa & 8M $\sim$ 15B & MLM \\
    \href{https://github.com/evolutionaryscale/esm}{ESM C}~\cite{esmc} & \gape{\makecell[c]{UniRef, \href{https://www.ebi.ac.uk/metagenomics}{MGnify}~\cite{richardson2023mgnify} \& \\ \href{https://img.jgi.doe.gov/m/}{JGI}~\cite{chen2023img}}} 
    & AA Seq. & Encoder
    % \gape{\makecell[c]{Encoder w/ Pre-LN,\\ RoPE \& SwiGLU}} 
    & 300M $\sim$ 6B & MLM \\
    \href{https://huggingface.co/Rostlab/prot_bert}{ProtBERT}~\cite{elnaggar2021prottrans} & \href{https://www.uniprot.org/uniref?query=*&facets=identity%3A1.0}{UniRef100}~\cite{suzek2015uniref}\textsuperscript{1} / \href{https://bfd.mmseqs.com/}{BFD}~\cite{steinegger2019protein}\textsuperscript{2} & AA Seq. & BERT~\cite{devlin2018bert} & 420M & MLM \\
    \href{https://huggingface.co/Rostlab/prot_albert}{ProtAlbert}~\cite{elnaggar2021prottrans} & UniRef100 & AA Seq. & Albert~\cite{lan2019albert} & 224M & MLM \\
    \href{https://huggingface.co/Rostlab/prot_electra_generator_bfd}{ProtElectra}~\cite{elnaggar2021prottrans} & UniRef100 & AA Seq. & Electra~\cite{clark2020electra} & 420M & MLM \\
    \href{https://github.com/yarongef/DistilProtBert}{DistilProtBert}~\cite{geffen2022distilprotbert} & UniRef50 & AA Seq. & ProtBERT & 230M & MLM \\
    PMLM~\cite{he2021pre} & UniRef50 & AA Seq. & RoBERTa & 87M $\sim$ 715M & \gape{\makecell[c]{MLM \&\\ Pairwise MLM}} \\
    \href{https://github.com/ISYSLAB-HUST/ProtFlash}{ProtFlash}~\cite{wang2023deciphering} & UniRef50 & AA Seq. & Encoder w/ Linear Attn. & 174M & MLM \\
    % \cmidrule{2-7}
    \href{https://github.com/amazon-science/LC-PLM}{LC-PLM}~\cite{wang2024long} & UniRef50 \& UniRef90 & AA Seq. & BiMamba-S & 130M $\sim$ 4B & MLM \\
    \href{https://github.com/bytedance/dplm}{DPLM}~\cite{wang2024diffusion} & UniRef50 & AA Seq. & ESM-2~\cite{lin2023evolutionary} & 150M $\sim$ 3B & \gape{\makecell[c]{MLM \& \\ Discrete Diffusion}} \\
    \href{https://github.com/genbio-ai/AIDO}{AIDO.Protein}~\cite{sun2024mixture} & \gape{\makecell[c]{UniRef90 \& \\ \href{https://colabfold.mmseqs.com/}{ColabFoldDB}~\cite{mirdita2022colabfold}\textsuperscript{3}}} & AA Seq. & Encoder w/ MoE~\cite{jiang2024mixtral} & 16B & MLM \\
    ProtTXL~\cite{elnaggar2021prottrans} & UniRef100 / BFD & AA Seq. & Transformer-XL~\cite{dai2019transformer} & 409M / 562M & NTP \\
    \href{https://huggingface.co/Rostlab/prot_xlnet}{ProtXLNet}~\cite{elnaggar2021prottrans} & UniRef100 & AA Seq. & XLNet~\cite{yang2019xlnet} & 409M & NTP \\
    \href{https://huggingface.co/nferruz/ProtGPT2}{ProtGPT2}~\cite{ferruz2022protgpt2} & UniRef50 & AA Seq. & GPT-2~\cite{radford2019language} & 738M & NTP
    \\
    \href{https://github.com/lightonai/RITA}{RITA}~\cite{hesslow2022rita} & UniRef100 & AA Seq. & GPT-3~\cite{brown2020language} & 85M $ \sim $ 1.2B & NTP \\
    \href{https://github.com/salesforce/progen/tree/main/progen2}{ProGen2}~\cite{nijkamp2023progen2} & \href{https://www.uniprot.org/uniref?query=*&facets=identity%3A0.9}{UniRef90}~\cite{suzek2015uniref}\textsuperscript{1} \& BFD30\textsuperscript{2} & AA Seq. & Decoder & 151M $ \sim $ 6.4B & NTP \\
    \href{https://huggingface.co/Rostlab/prot_t5_xl_UR50}{ProtT5}~\cite{elnaggar2021prottrans} & UniRef50 / BFD & AA Seq. & T5~\cite{raffel2020exploring} & 3B, 11B & MLM \\ 
    \href{https://github.com/agemagician/Ankh}{Ankh}~\cite{elnaggar2023ankh} & UniRef50 & AA Seq. & Encoder-Decoder & 755M, 1.9B & CSR \\
    \href{https://huggingface.co/proteinglm}{xTrimoPGLM}~\cite{chen2024xtrimopglm} & \gape{\makecell[c]{UniRef90 \& \\ ColabFoldDB}} & AA Seq. & GLM~\cite{du2021glm} & 100B & MLM \& CSR \\
    \href{https://github.com/jeffreyruffolo/AntiBERTy}{AntiBERTy}~\cite{ruffolo2021deciphering} & \href{https://opig.stats.ox.ac.uk/webapps/oas}{OAS}~\cite{kovaltsuk2018observed}\textsuperscript{4} & AA Seq. & BERT & 26M & MLM \\
    \href{https://github.com/alchemab/antiberta}{AntiBERTa}~\cite{leem2022deciphering}  & OAS \& \href{http://opig.stats.ox.ac.uk/webapps/sabdab}{SAbDab}~\cite{dunbar2014sabdab}\textsuperscript{5} & AA Seq. & RoBERTa & 86M & MLM \\
    \href{https://github.com/TobiasHeOl/AbLang}{AbLang}~\cite{olsen2022ablang} & OAS & AA Seq. & RoBERTa & 125M & MLM \\
    \href{https://github.com/KyGao/ABGNN}{AbBERT}~\cite{gao2023pre} & OAS & AA Seq. & BERT & 110M & Masked CDR Recon. \\
    \href{https://github.com/IBM/ReprogBERT}{ReprogBert}~\cite{melnyk2023reprogramming} & SAbDab & AA Seq. & \textcolor{navy}{BERT} w/ AA Emb. & 110M, 340 & Masked CDR Recon. \\
    \href{https://github.com/Graylab/IgLM}{IgLM}~\cite{shuai2023iglm} & OAS & \gape{\makecell[c]{ID Tag | AA Seq.}} & GPT-2 & 13M & CSR \\
    \href{https://github.com/oxpig/p-IgGen}{g-IgGen}~\cite{turnbull2024p} & OAS & AA Seq. & GPT-2 & 17M & NTP \\
    pAbT5~\cite{chu2023generative} & OAS & AA Seq. & \textcolor{pink}{ProtT5} & 3B & Light-Heavy Trans. \\
    \midrule
    \href{https://github.com/facebookresearch/esm}{ESM-MSA-1b}~\cite{rao2021msa} & UniRef50 & MSA & \gape{\makecell[c]{Encoder w/ Tied Row \\ \& Column Attn.}} & 100M & MLM \\
    MSA2Prot~\cite{ram2022few} & \href{https://pfam.xfam.org/}{Pfam}~\cite{mistry2021pfam} & MSA & \gape{\makecell[c]{Encoder w/ Axial Attn. - \\ Decoder w/ MSA Cross Attn.}} & - & NTP \\
    \href{https://github.com/Magiccircuit/MSA-Augmentor}{MSA-Augmenter}~\cite{zhang2023enhancing-msa}  & \gape{\makecell[c]{UniRef50 \& \\ \href{https://uniclust.mmseqs.com/}{UniClust30}~\cite{mirdita2017uniclust}\textsuperscript{7}}} & MSA & \gape{\makecell[c]{Encoder-Decoder w/ \\ Tied Row \& Column Attn.}} & 260M & NTP \\
    \href{https://github.com/THUDM/MSAGPT}{MSAGPT}~\cite{chen2024msagpt} & UniClust30 & \gape{\makecell[c]{Multi. AA Seq. + 2D PE}} & Decoder & 2.8B & NTP \\
    \href{https://github.com/OpenProteinAI/PoET}{PoET}~\cite{truong2024poet} & UniRef50 & Multi. AA Seq. & Decoder w/ Tiered Attn. & 400M & NTP \\
    \href{https://github.com/Bitbol-Lab/ProtMamba-ssm}{ProtMamba}~\cite{sgarbossa2024protmamba}  & \gape{\makecell[c]{\href{https://registry.opendata.aws/openfold/}{OpenProteinSet}~\cite{ahdritz2024openproteinset} \& \\ UniClust30}} & Multi. AA Seq. & Mamba~\cite{gu2023mamba} & 107M & CSR \\
    \href{https://huggingface.co/GleghornLab/cdsBERT}{cdsBERT}~\cite{hallee2023cdsbert}  & \href{https://www.ncbi.nlm.nih.gov/CCDS/CcdsBrowse.cgi}{CCDS}~\cite{pujar2018consensus} \& \href{https://asia.ensembl.org/index.html}{Ensembl} & \gape{\makecell[c]{AA Seq. \& Codon Seq.}} & \textcolor{pink}{ProtBERT}, \textcolor{navy}{Ankh} & 230M & MLM \& CL \\
    \href{https://huggingface.co/ChatterjeeLab/PTM-Mamba}{PTM-Mamba}~\cite{peng2024ptm} & \gape{\makecell[c]{UniProt \& \href{https://www.uniprot.org/uniprotkb?query=reviewed:true}{Swiss-Prot}~\cite{boutet2007uniprotkb}\textsuperscript{8}}} & \gape{\makecell[c]{AA Seq. \& PTM Seq.}} & Mamba~\cite{qu2024survey} & - & MLM \\
    \bottomrule
    \end{tabular}
    \begin{tablenotes}
    \vspace{0.1cm}
    \item[\textbf{*}] \textbf{Seq.} - Sequence; \textbf{PE} - Positional Encoding; \textbf{Encoder} - Transformer Encoder; \textbf{Decoder} - Transformer Decoder; \textbf{Attn.} - Attention; \textbf{Emb.} - Embeddings;\\ \textbf{MLM} - Masked Language Modeling; \textbf{NTP} - Next Token Prediction; \textbf{CRS} - Corrupted Spans Reconstruction; \textbf{Recon.} - Reconstruction; \textbf{Trans.} - Translation;\\ \textbf{CL} - Contrastive Learning; 
    \vspace{0.1cm}
     \item[1] UniRef database provides clustered sets of protein sequences based on the UniProt Knowledgebase (\href{https://www.uniprot.org/uniprotkb}{UniProtKB})~ and UniProt Archive (\href{https://www.uniprot.org/uniparc}{UniParc}) records. UniRef50, UniRef90, and UniRef100 are clusters of protein sequences at 50\%, 90\%, and 100\% identity. As of December 2024, they have approximately 68 million, 199 million, and 435 million results, respectively. 
     \setlength{\parskip}{0em}
     \item[2] Big Fantastic Database (BFD) combines UniProt with metagenomic data, containing over 2.6 billion protein sequences. BFD30 clusters the proteins at 30\% identity. 
     \item[3] ColabFoldDB is established by merging various metagenomic databases and contains over 740 million proteins.
     \item[4] Observed Antibody Space (OAS) collects and annotates over one billion antibody sequences from over 80 studies. 
     \item[5] Structural antibody database (SAbDab) provides antibody structures that are consistently annotated and presented.
     \item[6] As of December 2024, Pfam database collects approximately 23 thousand protein families, each represented by MSAs and hidden Markov models. 
     \item[7] UniClust databases cluster the UniProtKB sequences ased on 90\%, 50\%, and 30\% pairwise sequence identity levels. 
     \item[8] UniProtKB/Swiss-Prot contains manually reviewed textual annotations of proteins extracted from literature and analysis.
   \end{tablenotes}
    \end{threeparttable}
    \label{tab1}
\end{table*}

\subsection{Structure and Function Enhanced pLMs}

As sequence-based pLMs demonstrate the ability to capture implicit structural and functional semantics from protein sequences through large-scale pre-training, the further integration of explicit knowledge can enhance their understanding of proteins at a more comprehensive level. In this subsection, we present the recent advancements in constructing structure-and-function-enhanced pLMs. We explain the data form of protein structure and function individually, and introduce the corresponding incorporation methodologies. Relevant contents are also summarized in Table \ref{tab2}.

\subsubsection{Structure Enhanced pLMs}

\begin{figure*}[!t]%
  \centering
  \captionsetup{font=small}
  \includegraphics[width=0.9\linewidth]{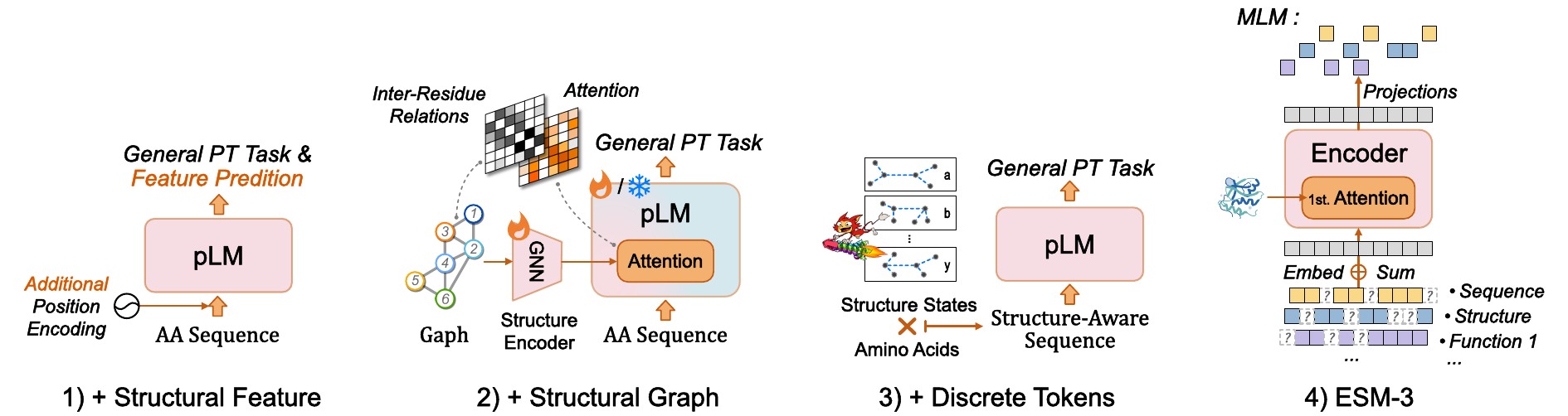}
  \caption{Typical \textbf{structure-enhanced pLMs}. \textbf{1)} Pre-calculated structural features can be injected into the input AA sequence as position encoding, or utilized in an additional training objective. \textbf{2)} Considering the significant correlation between the Transformer attention map and protein structural contacts, structural graphs can be encoded by GNNs and combined with the attention module of pLMs. \textbf{3)} Local structure states along the polypeptide chain are distilled into discrete tokens, which are subsequently involved in the training procedure of pLMs. \textbf{4)} ESM-3 presents the sequence, structure, and functions of protein as multiple tracks of discrete tokens, with all kinds of information fused within a unified latent space. In particular, there is an additional geometric attention contained in the first Transformer block to process the protein backbone structure.}
  \label{fig5}
\end{figure*}

As of December 2024, the Protein Data Bank (PDB)~\cite{wwpdb2019protein} has accumulated a vast repository of over 220 thousand experimentally determined protein structures. Furthermore, AlphaFold Protein Structure Database offers more than 200 million reliable structure prediction results. These invaluable resources establish explicit mapping relationships between sets of protein sequences and their 3D structures, serving as a robust data foundation for structure-enhanced pLMs. Figure \ref{fig5} presents three typical cases of integration of structural information through pre-processed structural features, the structural graph, or discrete tokens. 

Initially, each PDB file comprehensively documents the atom-scale 3D coordinates of a protein structure, and we can extract a wide variety of structural characteristics for more targeted learning. 
For example, PeTriBERT~\cite{dumortier2022petribert} incorporates additional position encodings for residues, including their central coordinates and rotation parameters. 
ESM-S~\cite{zhang2024structure} enhances ESM-2 by integrating a new training task, remote homology detection, which seeks to identify proteins with similar structures but low sequence similarity. 
METL~\cite{gelman2024biophysics} is trained to infer protein biophysical attributes, which are calculated from static structures through Rosetta~\cite{alford2017rosetta}.

Moreover, protein structures can be represented as graphs, where the nodes refer to amino acids, and the edges connect AAs that are close in distance. 
There is an important conjunction between the nature of graph and the principle of language models: graph represents relationships between nodes, while language models learn relationships between tokens. 
It is revealed that residue-residue contacts can be predicted from the attention maps of sequence-based pLMs~\cite{rao2020transformer}. 
Inversely, the structural graph can be incorporated into the attention map of pLMs in the construction of structure-enhanced pLMs. 
Additional GNN encoders (e.g., GVP~\cite{jing2020learning}, GearNet~\cite{zhang2022protein}) that represent structural inter-residue relations are integrated together with the attention mechanism of pLMs. For instance, PST~\cite{chen2024endowing} refines ESM-2 by attaching a novel structure extractor into each attention module, LM-Design~\cite{zheng2023structure} introduces a lightweight structural adapter into ESM-1b after the final Transformer layer, and ProseLM~\cite{ruffolo2024adapting}  introduce well-designed adapters that update the outputs of the attention \& feed-forward operations of each ProGen2 layer in fusing encoded structural features and low-rank language model embeddings.

In recent years, vector quantized-variational autoencoder (VQ-VAE)~\cite{van2017neural,qu2024tokenrec} has emerged as a flourishing technique that encodes the 3D protein structure into discrete tokens, representing the local geometric conformation of each amino acid. Technically speaking, VQ-VAE employs an encoder and a decoder to map data between the real and hidden space, and learns a codebook to quantize continuous latent representations into discrete ones. As the VQ-VAE tokenizer compresses intricate continuous data into a limited number of discrete latent representations, there is a basic consensus that a small codebook size usually leads to more information loss. 
In retrospect, Foldseek~\cite{van2024fast} pioneeringly leverages VQ-VAE to establish the 3D interaction (3Di) alphabet for protein structure. The maximally evolutionary conserved structure states are represented by twenty 3Di tokens, which facilitate fast and accurate protein structure searching. 
While Foldseek's 3Di tokens have demonstrated robust performance in protein retrieval, the coarse-grained nature of the small codebook still limits its structure reconstruction ability. 
To make up for this weakness, more protein structure tokenizers have been proposed, featuring inventive designs of the encoder, quantization, and decoder modules, along with the critical exploration of codebook size. 
ProTokens~\cite{lin2023protokens}, FoldToken series~\cite{gao2024foldtoken4}, and AIDO.st~\cite{meynard2024balancing} are all promising approaches to distill protein structure tokens, aiming to achieve the balance among varied factors, including the compression efficiency, retrieval ability, reconstruction ability, downstream usability, etc. 

With protein structure tokenization methods, each 3D structure can be described as an array of protein structure tokens, allowing for seamless integration with language models. 
For instance, 
ProstT5~\cite{heinzinger2023prostt5} is developed by incrementally training ProtT5 to translate between the structure 3Di tokens and the sequence AA tokens. 
SaProt~\cite{su2023saprot,su2024saprothub} leverages a Structure-Aware (SA) vocabulary that integrates AA tokens and 3Di tokens to effectively represent proteins in both perspectives of primary and tertiary structures. 
ProSST~\cite{li2024deprot} involves a well-designed sequence-structure disentangled attention to learn combined the relationship between protein AA sequences and protein structural token sequences. 
In protein representation learning benchmarks, those models excel in various downstream tasks~\cite{rao2019evaluating,xu2022peer}, especially in the zero-shot mutation effect prediction~\cite{notin2024proteingym}. 
Besides, DPLM-2~\cite{wang2024dplm} 
learns the joint protein sequence-structure distribution by applying discrete diffusion to the aligned AA and structure tokens. Therefore, DPLM-2 not only performs well in protein representation learning for predictive tasks but also skills in various generative scenarios, such as unconditional sequence-structure co-generation, structure-conditioned sequence generation, and so on.

Notably, ESM-3~\cite{hayes2025simulating} is another recent updation of the ESM series, where discrete protein semantic tokens are introduced as an essential concept. Unlike ESM-1b, ESM-2, and ESM C, which are solely learned from protein sequences, ESM-3 is designed to represent and reason over the sequence, structure, and function of proteins. The technical scheme is illustrated in Figure \ref{fig5}-4, ESM-3 is an all-to-all masked language model that synchronously conditions on and generates multiple separate tracks. 
At the input, the protein sequence is presented as AA tokens, the protein structure is presented as another track of discrete tokens meanwhile injected into the first transformer block, and then different aspects of protein function (e.g., solvent accessible surface area, function annotations) are presented as more token tracks. 
Subsequently, ESM-3 is trained with a special mask language modeling objective: masks are randomly sampled and applied to each track, and the masked tokens should be predicted at the output. Therefore, it learns a single latent space with all kinds of information fused. In evaluations, ESM-3 not only demonstrates excellent benchmarking results but also drives a case of real-world protein design. 
By using ESM-3 for protein generation, a new bright fluorescent protein is discovered. It is distant from known fluorescent proteins (with $\sim$58\% identity only) and should take 500 million years of evolution to emerge in the wild.

\subsubsection{Function Enhanced pLMs}

The sequence and structure of proteins are clear concepts denoting the linear arrangement and the spatial configuration of residues, respectively. In contrast, protein function represents a multifaceted notion that encompasses diverse categories. As shown in figure \ref{fig2}-F, protein function is labeled by experimental results or manual annotations, and can be documented in academic textual materials.

\begin{figure*}[!t]%
  \centering
  \includegraphics[width=0.9\linewidth]{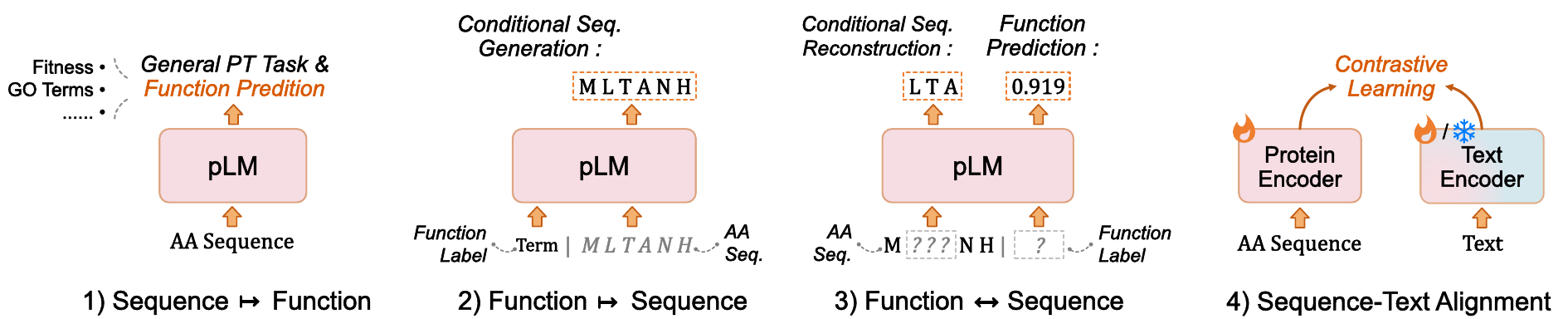}
  \captionsetup{font=small}
  \caption{Typical \textbf{function-enhanced pLMs}. \textbf{1-3)} pLMs learn the forward, inverse, and bidirectional correlations between protein sequences and functional labels. They undergo pre-training using various objectives, including function prediction, conditional sequence generation, or a variant of mask language modeling that combines the two.
  \textbf{4)} Protein sequences are aligned with their corresponding textual descriptions through contrastive learning.}
  \label{fig6}
\end{figure*}

To empower pLMs with the awareness of functional labels, existing studies guide pLMs to capture: 1) the forward correlation from protein sequences to functional labels, 2) the inverse correlation from functional labels to protein sequences, or 3) the bidirectional inter-correlations between them. 
Figure \ref{fig6} illustrates these three categories of methods. 

\begin{itemize}[leftmargin=8pt, itemsep=0pt, topsep=1pt, parsep=1pt]
\item First, function prediction objectives are introduced within pre-training frameworks. For instance, ProteinBERT~\cite{brandes2022proteinbert} combines mask language modeling with GO annotation prediction for pre-training. 
PromptProtein~\cite{wang2022multi} presents a multi-task pre-training framework tailored for pLMs. The PromptProtein model accepts an AA sequence as input, along with a prompt token that specifies the pre-training task, which can be MLM, $C_\alpha$ coordinate prediction, or protein-protein interaction prediction. 
Notably, ESM-1v~\cite{meier2021language} learns to score the sequence variation effect through a modified masked language modeling objective. With the mutated positions masked, ESM-1v is asked to compare the probability assigned to the mutant with that assigned to the wild-type. After such pre-training, ESM-1v can accurately capture the mutational effects on comprehensive protein function, even in the case of zero-shot inference. 
\item Second, some pLMs are trained to generate protein sequences conditioned on functional labels. For example, ProGen~\cite{madani2023large} enhances 280 million protein sequences from a wide range of families with prefix control tags that specify functional properties, and then proceeds the autoregressive pre-training. As a result, ProGen can generate functional proteins based on the provided control tag across diverse protein families. Similarly, ZymCTRL~\cite{munsamy2022zymctrl} is a conditional language model specifically pre-trained on the enzyme space, designed to generate enzyme sequences based on user-provided enzyme commission (EC) numbers.
\item Third, certain pLMs learn to perform protein function prediction and conditional sequence optimization collaboratively. 
While taking the concatenated functional property and AA sequence as input, Regression Transformer (RT)~\cite{born2023regression} undergoes unified pre-training that involves both masked property reconstruction and masked residue span reconstruction. Moreover, ProteinNPT~\cite{notin2024proteinnpt} integrates multiple sequence alignment (MSA) and auxiliary functional labels. ProteinNPT is a non-parametric Transformer with tri-axial self-attention across the residue/label tokens, the aligned homologous sequences, and the labeled instances, capable of both functional property prediction and iterative protein redesign. 
\end{itemize}

Text is a flexible and inclusive data modality that can encompass protein functional information. By aligning the representation space of protein sequences with their corresponding textual descriptions, pLMs acquire not only explicit protein function knowledge but also a certain level of natural language understanding. Figure \ref{fig6}-4 illustrates the typical bi-stream autoencoding framework. 
For instance, ProteinCLAP~\cite{liu2023text} performs contrastive learning (CL) to align protein-text modalities, thereby producing representations containing protein knowledge from text prompts. These representations could subsequently drive the text-guided protein design. 
Besides, ProtST~\cite{xu2023protst} is a multi-modal learning framework designed to enhance autoencoding pLMs. During pre-training, pLMs learn through the incorporated unimodal masked language modeling (MLM), cross-modal MLM, and protein-text CL. 
It is revealed that ProtST-induced pLMs outperform the vanilla pLMs in representation learning benchmarks.

Moreover, as one of the most successful contrastive learning methods, CLIP is first proposed to connect the text and images~\cite{radford2021learning}, training a text encoder and an image encoder to predict the pairings of a batch of (\textit{text}, \textit{image}) examples. In computational protein science, the technical framework of CLIP is employed to distill the same protein semantics described in different vocabularies. Based on the curated pairs of AA sequence and function annotation text, ProtET~\cite{yin2024multi} performs CLIP-like pre-training to align features of protein sequence and text, and ProteinCLIP~\cite{wu2024proteinclip} equips adapter layers for the pre-trained pLM and text encoder that each re-project the protein or text representations into a shared representation space. Furthermore, ProTrek~\cite{su2024protrek} is a tri-modal pLM that conducts contrastive learning of the sequence, structure, and function text. In addition to the comprehensive representation of proteins, ProtTrek supports all kinds of cross-and intra-modal retrieval with a total of nine searching tasks. We are allowed to precisely navigate the vast protein universe in seconds by giving the protein sequence, structure, or even natural language description. 

Besides, knowledge graphs (KGs) provide factual knowledge of protein functions that further integrate classification annotations and textual documentation. ProteinKG25~\cite{zhang2022ontoprotein} is a KG dataset that includes protein entities with sequences, Gene Ontology (GO) term entities with textual descriptions, as well as the GO-GO and protein-GO triples. On this basis, OntoProtein~\cite{zhang2022ontoprotein} performs incremental pre-training for ProtBERT through MLM and a knowledge embedding objective, and KeAP~\cite{zhou2022protein} performs protein-knowledge exploration at the token level with well-designed cross-attention modules. 
To compare those recent advances, Ko et al.~\cite{ko2024benchmarking} present a benchmark for functional text-integrated pLMs. In assessing the learned representations, the authors implement six models (ProteinCLAP, ProtST, ProteinCLIP, ProTrek, OntoProtein, and ESM-3) with a sequence-only baseline pLM (ESM-2) across six downstream tasks.
It is revealed that these function-enhanced pLMs outperform ESM-2 in five of six tasks, while no one is always the best.

Lastly, we simply review some relevant protein representation learning studies that aim for a comprehensive fusion of protein semantics. In integrating sequence and structure, 
SSEmb~\cite{blaabjerg2024ssemb} combines a pLM for multiple sequence alignments with a graph representation for the protein structure. ESM-GearNet~\cite{zhang2023systematic} explores different fusion strategies, including serial, parallel, and cross fusion, to combine sequence representations from pLMs (such as ESM-2) with structure representations from structure encoders (like GearNet). Lee et al.~\cite{lee2023pre} introduce a pre-training strategy that incorporates multi-view protein knowledge from the sequence, 3D structure, and surface for improved representation learning. 
BioCLIP~\cite{robinson2023contrasting} is a contrastive learning framework designed to train protein structure models by utilizing pLMs for assistance. 
Pursuing the unity of protein sequence-structure-function, MASSA~\cite{hu2023multimodal} aligns independently derived embeddings of AA sequence, structural graph, and GO terms. Additionally, Protein-Vec~\cite{hamamsy2023learning} is a multi-view information retrieval system for proteins, where a mixture-of-experts model is employed to combine protein sequences with seven structural and functional properties. 

\definecolor{navy}{RGB}{11, 11, 255}
\definecolor{pink}{RGB}{224, 33, 137}
\definecolor{gray}{RGB}{238, 238, 238}

\begin{table*}[!t]\scriptsize
    \centering
    \captionsetup{font=small}
    \caption{\textbf{Structure and Function Enhanced pLMs}. For each 
    pLM, the resource link, pre-training corpora, input data format, network architecture, size of full parameters, and pre-training objective is summarized here. Unusual abbreviations are explained in the footnote \textbf{*}, and important databases are briefly introduced in footnotes 1 to 9. In the "Architecture" field, models employed in a frozen form are colored \textcolor{navy}{blue}, models fine-tuned are colored \textcolor{pink}{pink}, and the others are trained from scratch.}
    \begin{threeparttable}[b]
    \rowcolors{2}{gray}{white}
    \begin{tabular}{ccccccc}
    \toprule
    \textbf{Method} & \textbf{Corpora} & \textbf{Input} & \textbf{Architecture} & \textbf{\#Parameter} & \textbf{Objective} \\
    \midrule
    PeTriBERT~\cite{dumortier2022petribert} & \href{https://alphafold.ebi.ac.uk/}{AlphaFoldDB}~\cite{varadi2022alphafold}\textsuperscript{1} & AA Seq. + Struct. PE & BERT & 40M & MLM \\
    \href{https://github.com/DeepGraphLearning/esm-s}{ESM-S}~\cite{zhang2024structure} & DeepSF~\cite{hou2018deepsf} & AA Seq. & \textcolor{pink}{ESM-2} & 8M $ \sim $ 650M & \gape{\makecell[c]{Remote Homology\\ Detection} }\\
    \href{https://github.com/gitter-lab/metl}{METL}~\cite{gelman2024biophysics} & - & AA Seq. + Struct. PE & Encoder & 20M, 50M & Biophysics Pred. \\
    \midrule
    \href{https://github.com/BytedProtein/ByProt}{LM-Design}~\cite{zheng2023structure} & \href{https://www.cathdb.info/}{CATH}~\cite{orengo1997cath}\textsuperscript{3} & AA Seq. \& Struct. Graph & \textcolor{navy}{ESM-1b} w/ Struct. Adapter & 658M  & Struct.-Cond. MLM \\
    \href{https://github.com/Profluent-AI/proseLM-public}{ProseLM}~\cite{ruffolo2024adapting} & CATH / \href{https://www.rcsb.org/}{PDB}~\cite{wwpdb2019protein}\textsuperscript{2} & AA Seq. \& Struct. Graph & \textcolor{navy}{ProGen2} w/ Struct. Adapter & 151M $\sim$ 6.4B & Struct.-Cond. NTP \\
    \href{https://github.com/BorgwardtLab/PST}{PST}~\cite{chen2024endowing} & AlphaFoldDB & AA Seq. \& Struct. Graph & \textcolor{navy}{ESM-2} w/ Struct. Extractors & 8M $ \sim $ 650M & MLM \\
    \midrule
    \href{https://github.com/mheinzinger/ProstT5}{ProstT5}~\cite{heinzinger2023prostt5} & AlphaFoldDB & AA Seq. \& 3Di Tokens & \textcolor{pink}{ProtT5} & 3B & CSR \& AA-3Di Trans. \\
    \href{https://github.com/westlake-repl/SaProt}{SaProt}~\cite{su2023saprot} & AlphaFoldDB \& PDB & Struct.-Aware Tokens & ESM-2 & 35M, 650M & MLM \\
    \href{https://github.com/ai4protein/ProSST}{ProSST}~\cite{li2024deprot} & AlphaFoldDB & AA Seq. \& Struct. Tokens & \gape{\makecell[c]{Encoder w/ \\Disentangled Attn.}} & 110M & MLM \\
    \href{https://bytedance.github.io/dplm/dplm-2}{DPLM-2}~\cite{wang2024dplm} & PDB \& AlphaFoldDB & AA Seq. \& Struct. Tokens & \textcolor{pink}{DPLM} & 150M $\sim$ 3B & Discrete Diffusion \\
    \href{https://github.com/evolutionaryscale/esm}{ESM-3}~\cite{hayes2025simulating} & \gape{\makecell[c]{UniRef, \href{https://www.ebi.ac.uk/metagenomics}{MGnify}~\cite{richardson2023mgnify},\\ \href{https://img.jgi.doe.gov/m/}{JGI}~\cite{chen2023img}, OAS, PDB,\\ AlphaFoldDB, \href{https://esmatlas.com/}{ESMAtlas}~\cite{lin2023evolutionary}\textsuperscript{4},\\ \href{https://www.ebi.ac.uk/interpro/}{InterPro}~\cite{paysan2023interpro}\textsuperscript{5},\\ \href{https://www.ebi.ac.uk/jdispatcher/pfa/iprscan5}{InterProScan}~\cite{jones2014interproscan}\textsuperscript{5}}} & \gape{\makecell[c]{AA Seq., Struct. Tokens \\ \& Func. Tokens}} & \gape{\makecell[c]{Encoder w/ \\ Struct. Block}} & 1.4B $\sim$ 98B & MLM \\
    \midrule
    \href{https://github.com/nadavbra/protein_bert}{ProteinBERT}~\cite{brandes2022proteinbert} & UniRef90 \& \href{https://geneontology.org/}{GO}~\cite{ashburner2000gene}\textsuperscript{6} & AA Seq, \& GO Annot. & Customized Encoder & 16M & NPT \& GO Pred. \\
    \href{https://github.com/HICAI-ZJU/PromptProtein}{PromptProtein}~\cite{wang2022multi} & \gape{\makecell[c]{UniRef50, PDB \&\\ \href{https://string-db.org/}{STRING}~\cite{szklarczyk2019string}\textsuperscript{7}}} & AA Seq. | Prompt Token & \gape{\makecell[c]{Encoder w/ \\Prompt-Aware Attn.}} & 650M & \gape{\makecell[c]{MLM, Struct. Pred. \\ \& PPI Pred.}} \\
    \href{https://github.com/facebookresearch/esm}{ESM-1v}~\cite{meier2021language} & UniRef90 & AA Seq. & ESM-1b & 650M & Mutation Scoring MLM \\
    \midrule
    \href{https://github.com/gitter-lab/metl}{ProGen}~\cite{madani2023large} & \gape{\makecell[c]{Pfam, GO, \& \\
    \href{https://www.ncbi.nlm.nih.gov/taxonomy}{NCBI Taxonomy}~\cite{federhen2012ncbi}}} & Func. Tag | AA Seq. & Decoder & 1.2B & Func.-Cond. NTP \\
    \href{https://huggingface.co/AI4PD/ZymCTRL}{ZymCTRL}~\cite{munsamy2022zymctrl} & \href{https://www.brenda-enzymes.org/}{BRENDA}\textsuperscript{8} & EC Number | AA Seq. & Constomized Decoder & 738M & Func.-Cond. NTP \\
    \href{https://github.com/GT4SD/gt4sd-core/tree/main/examples/regression_transformer}{RT}~\cite{born2023regression} & \href{https://github.com/songlab-cal/tape}{TAPE}~\cite{rao2019evaluating} & Prop. Label | AA Seq. & XLNet w/ Bidirec. Attn. & 27M & CSR \& Prop. Pred. \\
    \href{https://github.com/OATML-Markslab/ProteinNPT}{ProteinNPT}~\cite{notin2024proteinnpt} & - & MSA | Prop. Label & \gape{\makecell[c]{Encoder w/ Tied Row \\ \& Column Attn.}} & - & MLM \& Prop. Pred. \\
    \midrule
    \href{https://github.com/chao1224/ProteinDT}{ProteinCLAP}~\cite{liu2023text} & \href{https://www.uniprot.org/uniprotkb?query=reviewed:true}{Swiss-Prot}~\cite{boutet2007uniprotkb}\textsuperscript{9} & AA Seq. \& Func. Text & \textcolor{pink}{ProtBERT}, \textcolor{pink}{SciBERT}~\cite{beltagy2019scibert} & 420M & Seq.-Text CL \\
    ProtET~\cite{yin2024multi} & Swiss-Prot \& \href{https://www.uniprot.org/uniprotkb?query=reviewed:false}{TrEMBL}~\cite{boutet2007uniprotkb}\textsuperscript{9} & AA Seq. \& Func. Text & \textcolor{pink}{ESM-2}, \textcolor{pink}{PubMedBERT}~\cite{gu2021domain} & 750M & Seq.-Text CL \\
    \href{https://github.com/wukevin/proteinclip}{ProteinCLIP}~\cite{wu2024proteinclip} & UniProtKB & AA Seq. \& Func. Text & \textcolor{navy}{ESM-2} / \textcolor{navy}{ProtT5} & 8M $\sim$ 3B & Seq.-Text CL \\
    \href{https://huggingface.co/westlake-repl/ProTrek_650M_UniRef50}{ProTrek}~\cite{su2024protrek} & \gape{\makecell[c]{UniRef50, AlphaFoldDB,\\ PDB, \& Swiss-Prot}} & 
    \gape{\makecell[c]{AA Seq., Struct. Tokens \&\\ Func. Text}} & \gape{\makecell[c]{\textcolor{pink}{ESM-2}, Encoder, \\ \textcolor{pink}{PubMedBERT}}} & 930M & Seq.-Struct.-Func. CL \\
    \href{https://github.com/DeepGraphLearning/ProtST}{ProtST}~\cite{xu2023protst} & Swiss-Prot & AA Seq. \& Func. Text & \gape{\makecell[c]{\textcolor{pink}{ProtBERT} / \textcolor{pink}{ESM-1b} / \\ \textcolor{pink}{ESM-2}, \textcolor{navy}{PubMedBERT}}} & 420M / 650M & MLM \& CL \\
    \midrule
    \href{https://github.com/zjunlp/OntoProtein}{OntoProtein}~\cite{zhang2022ontoprotein} & \href{https://github.com/zjunlp/OntoProtein/}{ProteinKG25}~\cite{zhang2022ontoprotein} & AA Seq. \& KG & \textcolor{pink}{ProtBERT}, \textcolor{navy}{PubMedBERT} & 420M & KE \& MLM \\
    \href{https://github.com/RL4M/KeAP}{KeAP}~\cite{zhou2022protein} & ProteinKG25 & AA Seq. \& KG & \gape{\makecell[c]{Encoder-Decoder, \\ \textcolor{navy}{PubMedBERT}}} & - & MLM \\
    \bottomrule
    \end{tabular}
    \begin{tablenotes}
    \vspace{0.1cm}
    \item[\textbf{*}] \textbf{Struct.} - Structure; \textbf{Func.} - Function; \textbf{Prop.} - Functional Property; \textbf{Annot.} - Annotation; \textbf{KG} - Knowledge Graph; \textbf{Attn.} - Attention;\\ \textbf{Pred.} - Prediction; \textbf{Cond.} - Conditioned; \textbf{Trans.} - Translation; \textbf{PPI} - Protein-Protein Interaction; \textbf{KE} - Knowledge Embedding
    \vspace{0.1cm}
    \item[1] AlphaFold Database (AlphaFoldDB) provides reliable structure predictions for over 200 million proteins. 
    \item[2] Protein Data Bank (PDB) collects more than 220 thousand experimentally-determined 3D structures of proteins. 
    \item[3] CATH is a classification of protein structures based on PDB. 151 million protein domains are grouped into over 5 thousand superfamilies. 
    \item[4] ESM Metagenomic Atlas (ESMAtlas) contains 772 million metagenomic protein structures predicted by ESMFold. 
    \item[5] InterPro and InterProScan provide functional analysis of proteins by categorizing them into families and predicting their domains and crucial sites. 
    \item[6] Gene Ontology (GO) knowledgebase systematically annotates protein functions with GO terms. 
    \item[7] STRING integrates all public protein-protein interaction (PPI) sources, covering over 2 billion PPIs among 59.3 million proteins. 
    \item[8] BRENDA database contains approximately 37 million enzyme sequences and their corresponding EC number annotations. 
    \item[9] UniProtKB/Swiss-Prot provides manually reviewed protein annotations; UniProtKB/TrEMBL consists of proteins with computationally analyzed annotations. 
    \end{tablenotes}
    \end{threeparttable}
    \label{tab2}
\end{table*}

\subsection{Multimodal pLMs}

\begin{figure}[!t]%
  \centering
  \includegraphics[width=0.8\linewidth]{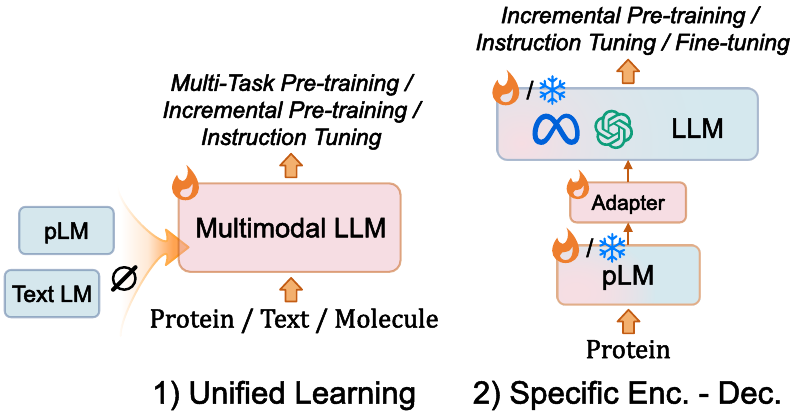}
  \captionsetup{font=small}
  \caption{Typical \textbf{multimodal pLMs}. \textbf{1)} In unified multimodal learning, multiple scientific languages share a unified latent space. We can perform multi-task pre-training from scratch, as well as incremental pre-training or instruction tuning based on pre-trained LMs. \textbf{2)} In "specific encoder - unified decoder" models, one or more specific encoders are connected to a unified decoder, with internal adapters typically learning to bridge the semantic spaces. The end-to-end model cam learn through incremental pre-training, instruction tuning, or fine-tuning.} 
  \label{fig7}
\end{figure}

In the preceding subsections, we have introduced existing pLMs that decipher protein sequences and understand the structural and functional information. While some of these models incorporate textual descriptions linked to proteins, their primary focus remains on protein-centered semantics. In this subsection, we introduce pLMs that exhibit proficiency in external languages, encompassing natural language with world knowledge, chemical molecule language, and beyond. As those languages convey greatly varied semantics, here we recognize them as different modalities. 
Those multimodal pLMs are summarized in Table \ref{tab3}. Meanwhile, Figure \ref{fig7} illustrates two typical technical approaches: unified multimodal learning, or "specific encoder - unified decoder" models. 

\textbf{Protein-Text Models} present the emergent ability of general LLMs and exhibit their superiority in instruction understanding and multi-task processing. InstructProtein~\cite{wang2023instructprotein} undergoes pre-training on both natural language and protein sequence corpora, followed by instruction tuning to align between these two distinct languages. ProtLLM~\cite{zhuo2024protllm} designs a protein-as-word language modeling approach, associating named protein entities in broad biological corpus with their AA sequences. This approach effectively unifies the protein and word inference as an autoregressive next-token prediction task. 
ProLLaMA~\cite{lv2024prollama} is a novel training framework that empowers general LLMs with the capability of protein language processing. Based on the low-rank adaptation (LoRA)\cite{hu2021lora} technique, this framework consists of two stages: incremental training using protein sequences and instruction tuning on the instruction dataset. 
Furthermore, Evolla~\cite{zhou2025decoding}, a multimodal pLM with 80 billion parameters, is developed for unraveling protein molecular mechanisms via natural language dialogue. To overcome the quantity and quality bottlenecks in the paired protein-text data, Evolla is trained on an unprecedented AI-generated dataset involving 546 million protein-question-answer triples, which could narrow the scale gap between protein sequences and the corresponding functional annotations. 

In biological systems, protein-molecule interactions play a pivotal role, which is especially important for medicine. This emphasizes the significance of bi-lingual \textbf{Protein-Molecule Models}. DrugGPT~\cite{li2023druggpt} is an autoregressive LM trained on a substantial amount of protein-ligand binding data. It can generate potential ligand SMILES conditions corresponding to specific protein sequences. ESM All-Atom (ESM-AA)~\cite{zheng2024multi} enables unified molecular modeling at both the residue scale for proteins and the atom scale for small molecules. In the development of ESM-AA, the main technical contributions include a multi-scale position encoding scheme that captures relationships among residues and atoms, and a pre-training framework for code-switching-enhanced protein sequences.

\definecolor{navy}{RGB}{11, 11, 255}
\definecolor{pink}{RGB}{224, 33, 137}
\definecolor{gray}{RGB}{238, 238, 238}

\begin{table*}[!t]\scriptsize
    \centering
    \captionsetup{font=small}
    \caption{\textbf{Multimodal pLMs}. In this table, we present the resource link, pre-training corpora, input data format, network architecture, scale of parameters, and learning procedure for typical \textit{protein-text}, \textit{protein-molecule}, \textit{protein-text-molecule}, and \textit{broader modal} models. In the context of "Corpora", important databases are briefly described in footnotes 1 to 5. In the "Architecture" field, parameters in uncolored models are trained from scratch, pre-trained parameters in \textcolor{pink}{pink} models are trainable, and pre-trained parameters in \textcolor{pink}{pink} models are frozen.}
    \begin{threeparttable}[b]
    \rowcolors{2}{gray}{white}
    \begin{tabular}{ccccccc}
    \toprule
    \textbf{Method} & \textbf{Corpora} & \textbf{Input} & \textbf{Architecture} & \textbf{\#Parameter} & \textbf{Learning Procedure} \\
    \midrule
    \href{https://protllm.github.io/project/}{ProtLLM}~\cite{zhuo2024protllm} & \gape{\makecell[c]{UniProt, \href{https://pubmed.ncbi.nlm.nih.gov/}{PubMed}~\cite{canese2013pubmed}\textsuperscript{1}, STRING \&\\ \href{https://github.com/zjunlp/Mol-Instructions}{Mol-Instructions}~\cite{fang2023mol}}} & AA Seq. \& Text & \gape{\makecell[c]{\textcolor{pink}{ProtST},\\ \textcolor{pink}{LLaMA-2}~\cite{touvron2023llama}}} & 7B & \gape{\makecell[c]{Protein-as-word\\ Pre-training}} \\
    \href{https://github.com/Lyu6PosHao/ProLLaMA}{ProLLaMA}~\cite{lv2024prollama} & UniRef50 \& InterPro & AA Seq. \& Text & \textcolor{pink}{LLaMA-2} & 7B & Instruction Tuning \\
    \href{https://github.com/HICAI-ZJU/InstructProtein}{InstructProtein}~\cite{wang2023instructprotein} & UniRef100 \& PubMed & AA Seq. \& Text & Decoder & 1.3B & \gape{\makecell[c]{Multimodal Pre-training \&\\ Instruction Tuning}} \\
    \href{http://www.chat-protein.com/}{Evolla}~\cite{zhou2025decoding} & \gape{\makecell[c]{Swiss-Prot, ProTrek~\cite{su2024protrek},\\ TrEMBL \& AI-generated data}} & AA Seq. \& Text & \gape{\makecell[c]{\textcolor{navy}{SaProt},\\ \textcolor{navy}{LLaMA-3}~\cite{dubey2024llama},\\ Align Module}} & 10B, 80B & \gape{\makecell[c]{Causal Language Modeling \&\\ Direct Preference Optimization}} \\
    \midrule
    \href{https://github.com/LIYUESEN/druggpt}{DrugGPT}~\cite{li2023druggpt} & \href{http://zinc20.docking.org/}{ZINC20}~\cite{irwin2020zinc20}\textsuperscript{2} \& \href{https://huggingface.co/datasets/jglaser/binding_affinity}{Jglaser}\textsuperscript{3} & AA Seq. \& SMILES & GPT-2 & 1.5B & \gape{\makecell[c]{Ligand \& Protein-Ligand\\ Pre-training}} \\
    \href{https://github.com/zhengkangjie/ESM-AA}{ESM-AA}~\cite{zheng2024multi} & AlphaFoldDB \& Uni-Mol~\cite{zhou2023uni} & \gape{\makecell[c]{AA Seq. \& SMILES \\ + Multi-scale PE}} & \textcolor{pink}{ESM-2} & 35M & \gape{\makecell[c]{Pre-training w/ MLM \&\\ Pairwise distance recovery}} \\
    \midrule
    \href{https://github.com/PharMolix/OpenBioMed}{BioMedGPT}~\cite{luo2023biomedgpt} & \gape{\makecell[c]{UniProt \& \href{https://pubchem.ncbi.nlm.nih.gov/}{PubChem}~\cite{kim2023pubchem}\textsuperscript{4}}} & \gape{\makecell[c]{AA Seq., Text \& \\ Mol. Graph}} & \gape{\makecell[c]{\textcolor{pink}{GraphMVP}~\cite{liu2021pre},\\ \textcolor{pink}{ESM-2}, \textcolor{pink}{LLaMA-2}}} & 10B & Multimodal Fine-tuning \\
    \href{https://github.com/QizhiPei/BioT5}{BioT5}~\cite{pei2023biot5} & \gape{\makecell[c]{UniRef50, ZINC20, PubChem,\\ PubMed, \href{https://www.tensorflow.org/datasets/catalog/c4}{C4}~\cite{raffel2020exploring}\textsuperscript{5} \& Swiss-Prot}} & \gape{\makecell[c]{AA Seq., Text \& \\ SELFIES}} & T5 & 252M & \gape{\makecell[c]{Pre-training w/ CSR \&\\ Multimodal Translation}} \\
    \href{https://github.com/QizhiPei/BioT5}{BioT5+}~\cite{pei2024biot5+} & \gape{\makecell[c]{UniRef50, ZINC20, PubChem,\\ PubMed, C4 \& Swiss-Prot}} & \gape{\makecell[c]{AA Seq., SELFIES, \\ IUPAC \& Text}} & T5 & 252M & \gape{\makecell[c]{Pre-training w/ CSR \&\\ Multimodal Translation}} \\
    InstructBioMol~\cite{zhuang2024instructbiomol} & \gape{\makecell[c]{PubMed, \href{https://www.biorxiv.org/}{bioRxiv}, \href{https://chemrxiv.org/}{ChemRxiv},\\ PubChem, UniRef50, \href{https://github.com/blender-nlp/MolT5/tree/main/ChEBI-20_data}{ChEBI-20}~\cite{edwards2022translation},\\ TrEMBL, Swiss-Prot,\\ \href{https://www.bindingdb.org/}{BindingDB}~\cite{gilson2016bindingdb} \& \href{https://www.rhea-db.org/}{Rhea}~\cite{bansal2022rhea}}} & \gape{\makecell[c]{SELFIES,\\ Mol Graph, AA Seq.,\\ Struct. Graph \& Text}} & \gape{\makecell[c]{Struct. GNNs,\\ \textcolor{navy}{ESM-2}, \textcolor{navy}{SaProt},\\ \textcolor{pink}{LLaMA-2}}} & 6.8B & \gape{\makecell[c]{Continual Pre-training \&\\ Instruction Tuning}} \\
    \midrule
    \href{https://github.com/HanwenXuTHU/BioTranslatorProject}{BioTranslator}~\cite{xu2023multilingual} & \gape{\makecell[c]{STRING, ChEBI-20 \&\\ \href{https://hpo.jax.org/}{Human Phenotype Ontology}~\cite{kohler2021human}}} & \gape{\makecell[c]{Text, AA Seq.,\\ Gene Expression \&\\ Phenotype Pathway}} & \gape{\makecell[c]{\textcolor{pink}{PubMedBERT},\\ Non-Text Projectors}} & 100M & Contrastive Learning \\
    \href{https://galactica.org/}{Galactica}~\cite{taylor2022galactica} & - & \gape{\makecell[c]{AA Seq., SMILES,\\ DNA Seq., \& Code}} & Decoder & 120B & Prompt Pre-training \\
    \bottomrule
    \end{tabular}
    \begin{tablenotes}
    \vspace{0.1cm}
    \item[1] PubMed database contains the abstract and citation information for more than 37 million biomedical literature. 
    \item[2] ZINC20 is a chemical database that contains billions of molecules and supports precious searching. 
    \item[3] Jglaser dataset contains 1.9 million unique pairs of protein sequence \& ligand SMILES with experimentally determined binding affinities. 
    \item[4] PubChem provides information about chemical molecules, such as the SMILES string and IUPAC names. 
    \item[5] Colossal Clean Crawled Corpus (C4) is a collection of general English-language text sourced from the public web scrape. 
    \end{tablenotes}
    \end{threeparttable}
    \label{tab3}
\end{table*}

Furthermore, \textbf{Protein-Text-Molecule Models} aim to decipher multiple biomolecular languages in a consistent manner. BioMedGPT~\cite{luo2023biomedgpt} aligns the feature spaces of an autoencoding pLM, a molecular graph encoder, and an autoregressive LLM by using a question-answering-based fine-tuning approach. BioT5~\cite{pei2023biot5} is a pre-training framework proposed for comprehensive multi-language integration. The BioT5 model is trained across data of protein sequences, molecule SELFIES strings, scientific texts, and wrapped sentences using the corrupted span reconstruction and bidirectional translation objectives. Then, BioT5+~\cite{pei2024biot5+} incorporates IUPAC names to enhance molecular understanding, aiming to bridge the gap between specialized molecular representations and the corresponding textual descriptions. 
Notably, Mol-Instructions~\cite{fang2023mol} is a meticulously curated dataset that consists of protein-oriented, molecule-oriented, and biomolecular text instructions. This dataset facilitates the adaptation of general LLMs, such as the LLaMA series, by enabling effective instruction tuning to address diverse biomolecular tasks. 
Moreover, InstructBioMol~\cite{zhuang2024instructbiomol} is a novel LLM learned through a comprehensive any-to-any alignment of natural language, molecules, and proteins. InstructBioMol has exhibited critical capabilities of biomolecular instruction following and multimodal data understanding, and demonstrated the potential to serve as a digital research assistant in supporting practical biomolecular tasks.

Besides, there have been extensive explorations for broader scientific languages, as well as investigations into the methods for bridging multiple languages. For instance, BioTranslator~\cite{xu2023multilingual} learns a cross-modal translation to bridge the user-written text and the corresponding non-text biological data, such as protein sequences, gene expression vectors, and phenotype pathways. Galactica~\cite{taylor2022galactica} is an LLM capable of storing, combining, and reasoning about scientific knowledge. It demonstrates proficiency in handling diverse data modalities, including general text, LATEX code, programming code, molecular SMILES, protein sequences, and DNA sequences. 
Notably, BioBridge~\cite{wang2023biobridge} is a parameter-efficient learning framework that bridges independently pre-trained scientific LLMs based on the biomedical knowledge graph. In constructing multimodal scientific LLMs, BioBridege presents promising to overcome the limitations imposed by data scarcity and computational costs.

\section{Utilization and Adaptation of Protein Language Models}

While the presence of progressive information flow between protein sequence, structure, and function is widely acknowledged, the intricate principles underlying protein folding and functioning are still awaiting revelation. Structure prediction, function prediction, and protein design have long been fundamental modeling problems in the field of computational protein science. 
As pLMs possess a fundamental understanding of proteins, they have been utilized or adapted to these problems, yielding significant advancements and impacts. 

\subsection{Protein Structure Prediction}

\begin{figure*}[!t]%
  \centering
  \includegraphics[width=0.9\linewidth]{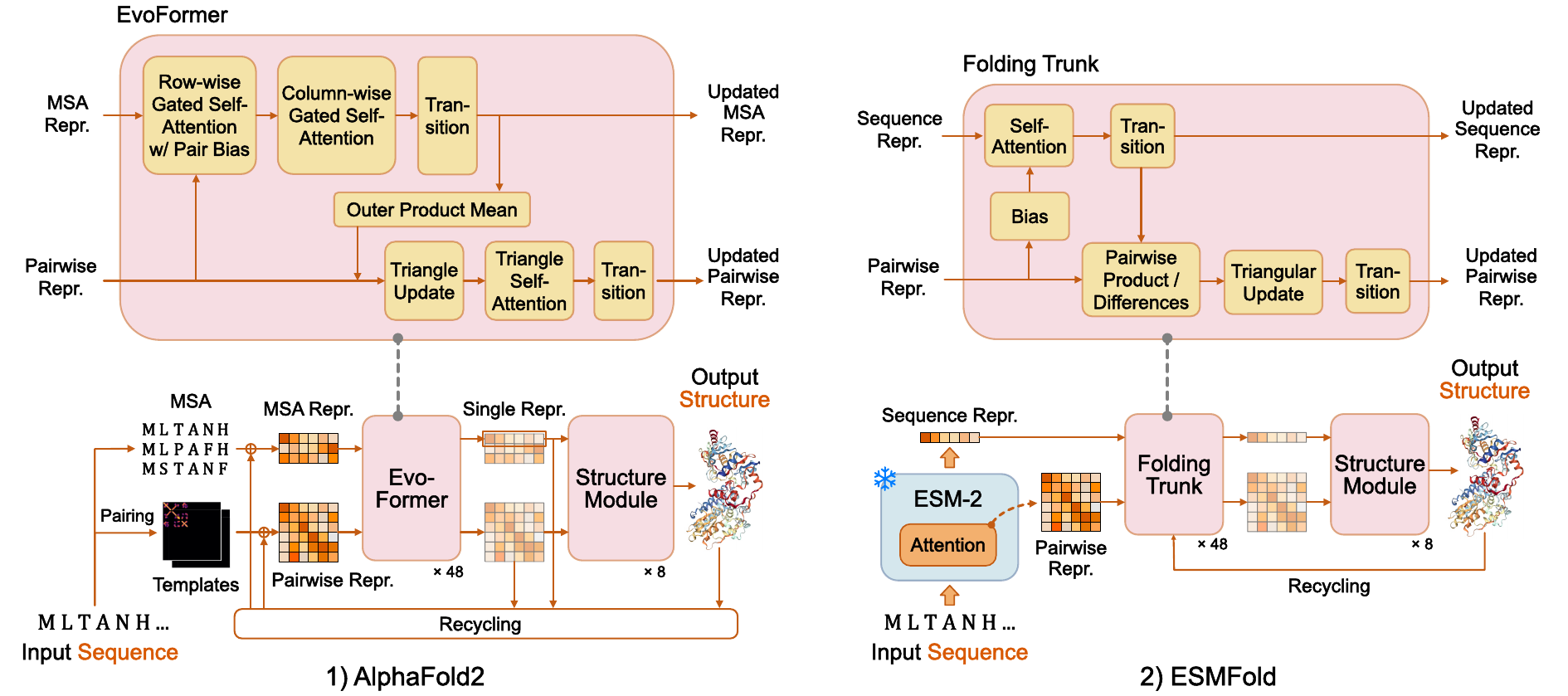}
  \captionsetup{font=small}
  \caption{\textbf{Workfolw overview of AlphaFold2}~\cite{jumper2021highly} \textbf{and ESMFold}~\cite{lin2023evolutionary}. 
  Both AlphaFold2 and ESMFold infer the high-resolution protein structure from a comprehensive understanding of the sequence. AlphaFold2 relies on MSA to gain evolutionary insights encoded in protein sequences, while ESMFold achieves this through the utilization of a protein language model. 
  }
  \label{fig8}
\end{figure*}

In structural biology, scientists determine protein structures through experimental techniques like X-ray crystallography~\cite{read2011new}, nuclear magnetic resonance~\cite{montelione2013recommendations}, and electron cryomicroscopy~\cite{henderson2012outcome}. These laboratory works are complex, time-consuming, and expensive. It is estimated that determining each protein structure takes months to years and costs tens of thousands of dollars~\cite{wei2019protein}. So far, there are only about two hundred thousand experimentally determined structures collected in the Protein Data Bank~\cite{wwpdb2019protein}. At this rate of development, it would take millions of research-years to analyze hundreds of millions of natural proteins that are sequenced but of unknown structures. In this context, protein structure prediction emerges as a crucial challenge. If a computational model can \textit{accurately infer the atom-wise 3D structures of proteins from their amino-acid sequences}, the progress of human understanding of protein structures would be significantly accelerated.

In recent years, rapid developments in artificial intelligence and computing power have greatly promoted the advancement of protein structure prediction. Breakthrough methods like AlphaFold2~\cite{jumper2021highly} and RoseTTAFold~\cite{baek2021accurate} exhibit an unprecedented level of near-experimental accuracy in predicting protein structures. They have played the role of essential tools for scientists to obtain a reliable protein structure within tens of minutes. The great success of these models should be credited to the ingenious incorporation of network architectures and training strategies, which well meet the evolutionary and geometric constraints on protein structures. Elaborate workflows are proposed to extract co-evolution knowledge from multiple sequence alignments (MSAs) and enforce the pairwise description of residues to satisfy the triangle inequality on distances.

For instance, Figure \ref{fig8}-1 illustrates the workflow of AlphaFold2. Given an input amino-acid sequence, AlphaFold2 first retrieves an MSA and a small number of homologous structures (i.e., templates). Notably, experience suggests that the MSA would significantly impact the model's performance more than the templates do. 
After initial encoding, the inference begins with an MSA representation that conveys evolutionary constraints and a pairwise representation that embodies geometric information. 
In the following Evoformer blocks, the two representations undergo constraint transformations and are updated mutually, which enables sufficient reasoning and fusion about the evolutionary and geometric information. Subsequently, all-atom positions and side-chain angles are inferred by a structure module. Besides, the model undergoes "recycling" iterations: the training loss is applied to outputs repeatedly, while the outputs are input back into the same modules recursively. 
Overall, within the latent space of AlphaFold2, the 3D structure of a protein is first hypothesized by an MSA and then progressively refined into an increasingly accurate prediction result.

Although these methods have demonstrated their effectiveness in protein structure prediction, they still encounter challenges due to the reliance on MSAs. First, searching for MSA from extensive databases is time-consuming, usually taking up most of the total inference time. It should be certain that speeding up protein structure prediction would further broaden its applications. Second, it's inherently difficult to obtain sufficient MSAs for orphan or fast-evolving proteins that lack homology. The reliance on MSA impedes the accuracy of structure prediction for these special proteins. Third, from a theoretical point of view, protein folding is an independent process for each protein. All structural information of a protein is fully encoded in its single sequence rather than the retrieved MSA. Models like AlphaFold2 learn MSA-to-structure but not sequence-to-structure, which deviates from the ultimate goal of grasping the rule of protein folding.

To overcome these MSA-caused challenges, methods for single-sequence (MSA-free) protein structure prediction have been developed, with pLMs playing a foundational role. Through large-scale pre-training, pLMs have learned the favorable patterns of protein sequence and some implicit knowledge of protein structure. When given a single amino-acid sequence as input, pLMs encode evolutionary constraints in sequence representations and embody the inter-residue structural relationships in attention maps (i.e., pairwise representations). In conveying the same information, the traditional MSA can be replaced by the sequence and pairwise representations generated by pLMs. 
This supports subsequent deep-learning modules to model protein folding as well.
Thus, from an end-to-end view, the typical workflow of single-sequence protein structure prediction involves taking pLMs as the foundation model and training additional "prediction heads" to deduce geometric constraints and complete structure predictions.

A series of methods build AlphaFold2-like "prediction heads" comprising a geometric module (i.e., EvoFormer) and a structure module, with the geometric module as the focus of improvement.
For example, the workflow of ESMFold~\cite{lin2023evolutionary} is presented in Figure \ref{fig8}-2. During inference, the sequence of a protein is fed into the foundation model (ESM-2), and the internal representations are extracted and sent to the folding trunk that contains an array of folding blocks. Resembling the EvoFormer block of AlphaFold2, each ESMFold's folding block updates the sequence representation and the pairwise representation alternately. Outputs of the folding trunk are then directed to an equivariant structure module, undergoing three rounds of recycling until the final protein structure prediction is produced.  
So far, ESMFold has been applied to predict structures for more than 617 million protein sequences, which provides a fire-new view into the diversity of natural proteins at the evolutionary scale. 
Furthermore, xT-Fold~\cite{chen2024xtrimopglm} has a similar architecture to ESMFold, while the number of stacked folding blocks is reduced from 48 to 1. 
HelixFold-Single~\cite{fang2023method} revises the Evoformer of AlphaFold2 into an EvoformerS (Evoformer with Single representations) module, where the column-wise gated self-attention originally designed for MSA is removed. 
OmegaFold~\cite{wu2022high} introduces Geoformer, a novel transformer neural network inspired by geometry principles. In Geoformer layers, the sequence and pairwise representations are iteratively smoothed, and the geometric inconsistency among them is gradually reduced. 

In addition, certain methods design distinct "prediction heads" that involve geometric modeling beyond updating the sequence and pairwise representations alternatively. 
In trRosettaX-Single~\cite{wang2022single}, the sequence and pairwise representations are concatenated and passed to Res2Net-Single, a multi-scale neural network that distills inter-residue 2D geometry. The predicted inter-residue distance and orientations are then utilized to reconstruct 3D structures through energy minimization. 
In RGN2~\cite{chowdhury2022single}, a pLM is combined with a Recurrent Geometric Network that uses Frenet–Serret frames to generate protein backbone structures. Furthermore, IgFold~\cite{ruffolo2023fast} is a specific antibody structure prediction model. IgFold consists of an antibody LM (i.e., AntiBERTy), followed by a series of customized graph networks and transformer layers that predict the backbone atom coordinates directly.

Compared to mainstream MSA-based methods, these pLM-based single-sequence structure prediction methods exhibit the advantages of being fast and universal. 
On the one hand, these methods are compared to AlphaFold2 and RoseTTAFold in terms of the inference runtime. For big proteins made up of several hundreds of amino acids, single-sequence protein structure prediction often cuts the runtime to less than one-third. While for short sequences of dozens of amino acids, there can be a hundredfold speed advantage. 
On the other hand, the structure prediction performance for proteins with no homology is also improved. In investigations, OmegaFold, trRosettaX-Single, and RGN2 are all observed to outperform AlphaFold2 and RoseTTAFold on those orphan proteins and \textit{de novo} designed proteins. However, there still remains room for improvement in accuracy: single-sequence protein structure prediction achieves competitive accuracy but is not superior to the MSA-to-structure prediction. Such a result indicates that the principle of protein folding is still not sufficiently grasped.

Recently, AlphaFold3~\cite{abramson2024accurate} has been proposed as the latest updation of the AlphaFold series, making surges in the field of protein science again. 
In addition to predicting protein structures solely, AlphaFold3 can infer the joint 3D structure of complexes, covering nucleic acids, small molecules, ions, and modified residues, and has demonstrated substantially improved accuracy over previous state-of-the-art specialized docking or interaction prediction tools. 
Technically, the advancement is achieved by the simpler but more general network architecture and training procedure. The reliance on MSA is significantly reduced by replacing the Evoformer of AlphaFold2 with a novel Pariformer module, where the MSA representation is removed, and all information is passed through the pairwise representation for geometric modeling. Moreover, the structure module of AlphaFold2 is replaced by a diffusion module to predict the raw atomic coordinates directly. Compared to AlphaFold3, pLM-based single-sequence protein structure prediction is advancing in a relatively consistent direction to stride the traditional MSA processing. However, there is still a long way to go toward understanding the fundamental physical and chemical laws of molecules. 

\subsection{Protein Function Prediction}

\begin{figure*}[!t]%
  \centering
  \includegraphics[width=0.85\linewidth]{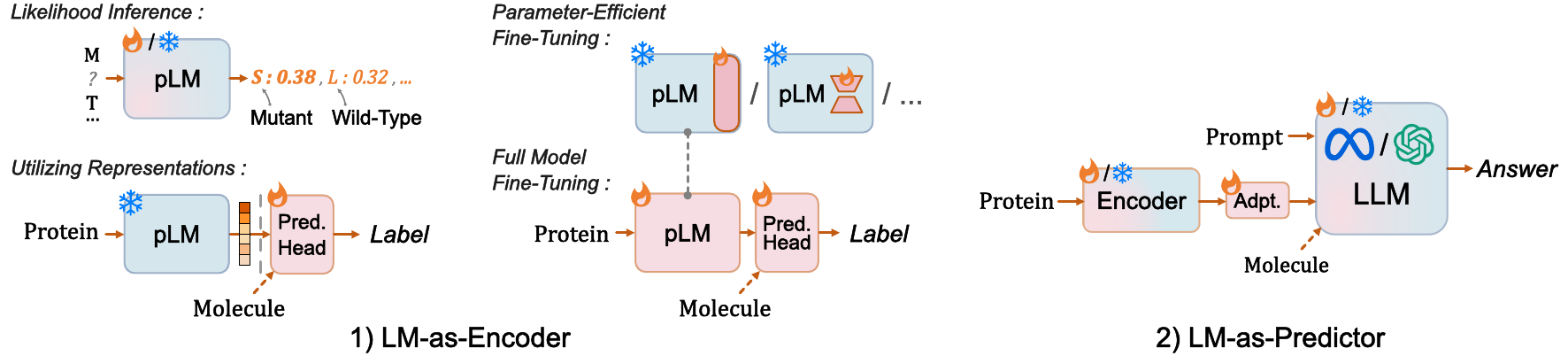}
  \captionsetup{font=small}
  \caption{\textbf{Typical technical schemes in pLM-based protein function prediction}. 
  1) In the "LM-as-Encoder" scheme, protein language models play a central role in the full models as encoders. Predictions could be inferred from the likelihood of language modeling or obtained through additional prediction heads. 
  2) In the "LM-as-Encoder" scheme, textual LLMs play a central role in the full models as predictors. In addition to the textual prompts, LLMs receive the encoded protein sequence/structure or molecule knowledge. 
  }
  \label{fig9}
\end{figure*}

Unlike the clearly defined protein sequence and structure, protein function exhibits multifaceted characteristics as different proteins play diverse biological roles in broad living systems. Then, greater gaps exist between the number of sequenced or structure-determined proteins and ground-truth function labels. In the Swiss-Prot~\cite{uniprot2023uniprot} database, less than 100 thousand proteins are manually annotated with Gene Ontology (GO) terms. For specific functional properties, such as stability, fluorescence, and fitness values, typical benchmark datasets (e.g., TAPE~\cite{rao2019evaluating}, FLIP~\cite{dallago2021flip}, PEER~\cite{xu2022peer}, ProteinGym~\cite{notin2024proteingym}) consist of roughly tens of thousands of labeled instances. Certain instances are under extreme data scarcity as the wet lab experiments have only been conducted on a minimal scale. As a result, protein function prediction has presented significant value in guiding the study of proteins that lack corresponding functional knowledge, where a wide range of different prediction tasks are involved.

Before the emergence of pLMs, AI models are individually trained from scratch for various protein function prediction tasks. This traditional paradigm had a tough drawback: as the models lack transferrable protein knowledge, the prediction performances are barely satisfactory, especially in the case of data scarcity. To overcome the shortcoming, pLMs have been successfully utilized or adopted in protein function prediction. Figure \ref{fig9} illustrates the typical utilization or adaptation techniques. In the "\textit{LM-as-Encoder}" scheme, prediction heads are typically incorporated following the main body of autoencoding pLMs. These heads are either trained independently with pLM parameters frozen or fine-tuned alongside the pLM in full-model or parameter-efficient manners. 
Even more simply, the likelihood probability derived in language modeling is skillfully transferred into predictions. In the "\textit{LM-as-Predictor}" scheme, a unified LLM generates answers according to prompts specifying contexts and questions. 
Consequently, the acquired knowledge within pLMs is effectively transferred to downstream predictions, resulting in significant advancements across various specific protein function prediction tasks. We organize this subsection by summarizing these tasks into five categories: functional property prediction, functional class annotation, functional site identification, protein-molecule interaction prediction, and multi-task question answering. Meanwhile, Table \ref{tab5} presents a comprehensive summary of relevant studies. 

\definecolor{navy}{RGB}{11, 11, 255}
\definecolor{pink}{RGB}{224, 33, 137}
\definecolor{gray}{RGB}{238, 238, 238}

\begin{table*}[t]\scriptsize
    \centering
    \captionsetup{font=small}
    \caption{\textbf{pLM-based Protein Function Prediction Methods}. Corresponding to the technical schemes summarized in Figure \ref{fig9}, we present the task category, resource link, encoder module, predictor module, and the employed technical scheme for each method. Unusual abbreviations are explained in the footnote \textbf{*}.}
    \begin{threeparttable}[b]
    \begin{tabular}{cccccc}
    \toprule
    \textbf{Sub Category} & \textbf{Method} & \textbf{Encoder} & \textbf{Predictor} & \textbf{Scheme} \\
    \toprule
    \multirow{18}{*}{\makecell[c]{Functional \\ Property \\ Prediction}} & \href{https://github.com/biomed-AI/LMDisorder}{LMDisorder}~\cite{song2023fast} & ProtT5 & Transformer-based Model & LM-as-Encoder (Repr.) \\
    & \href{https://github.com/SimonKitSangChu/EsmTherm}{ESMtherm}~\cite{chu2023protein} & ESM-2 & Linear Head & LM-as-Encoder (FM FT) \\
    & VISH-Pred~\cite{mall2024vish} & ESM-2 & Ensemble Classifier & LM-as-Encoder (FM FT) \\
    & \href{https://github.com/KangBoming/PIC}{PIC}~\cite{kang2024comprehensive} & ESM-2 & Customized Attention Model & LM-as-Encoder (FM FT) \\
    & \href{https://github.com/ChakradharG/PeptideBERT}{PeptideBERT}~\cite{guntuboina2023peptidebert} & ProtBERT & MLP Head & LM-as-Encoder (FM FT) \\
    & \href{https://github.com/ShuklaGroup/LassoESM}{LassoESM}~\cite{mi2024lassoesm} & ESM-2 & Customized Attention Model & LM-as-Encoder (FM FT) \\
    & \href{https://github.com/i-Molecule/optimalPh}{OphPred}~\cite{zaretckii2024approaching} & ESM-2 & XGBoost / KNN Classifier & LM-as-Encoder (Repr.) \\
    & \href{https://github.com/beckham-lab/EpHod}{EpHod}~\cite{gado2023deep} & ESM-1v & Customized Attention Model & LM-as-Encoder (Repr.) \\
    & \href{https://github.com/OATML-Markslab/ProteinGym}{ProteinGym}~\cite{notin2024proteingym} & \gape{\makecell[c]{ESM-1b / ESM-2 / ProtGPT2 /\\ ESM-MSA-1b / SaProt / ...}} & - & LM-as-Encoder (Infer.) \\
    & Brandes et al.~\cite{brandes2023genome} & ESM-1b & - & LM-as-Encoder (Infer.) \\
    & \href{https://github.com/ai4protein/Pro-FSFP}{FSFP}~\cite{zhou2024enhancing} & ESM-1v / ESM-2 / SaProt & - & LM-as-Encoder (PE FT \& Infer.) \\
    & \href{https://github.com/HICAI-ZJU/DePLM}{DePLM}~\cite{wangdeplm} & ESM-1v / ESM-2 & Denoising Module & LM-as-Encoder (PE FT \& Infer.) \\
    \midrule
    \multirow{8}{*}{\makecell[c]{Functional \\ Class \\ Annotation}} 
    & \href{https://github.com/kellylab/viral-protein-function-annotation-with-protein-language-model}{Flamholz et al.}~\cite{flamholz2024large} & ProtBERT & MLP Head & LM-as-Encoder (Repr.) \\
    & \href{https://github.com/ProteinEngineering-PESB2/ECNumberClassModels}{Fernandez et al.}~\cite{fernandez2023exploring} & ESM-1b & KNN / SVM / CNN Classifier & LM-as-Encoder (Repr.) \\
    & \href{https://github.com/biomed-AI/GraphEC}{GraphEC}~\cite{song2024accurately} & ProtT5 & \gape{\makecell[c]{GNN-based Model \&\\ Label Diffusion Algorithm}} & LM-as-Encoder (Repr.) \\
    & \href{https://dmiip.sjtu.edu.cn/ng3.0}{NetGO 3.0}~\cite{wang2023netgo} & ESM-1b & Logistic Regression Head & LM-as-Encoder (Repr.) \\
    & \href{https://github.com/bio-ontology-research-group/deepgo2}{DeepGO-SE}~\cite{kulmanov2024protein} & ESM-2 & Approximate Models & LM-as-Encoder (Repr.) \\ 
    & \href{https://github.com/biomed-AI/SPROF-GO}{SPROF-GO}~\cite{yuan2023fast} & ProtT5 & Label Diffusion Algorithm & LM-as-Encoder (Repr.) \\
    \midrule
    \multirow{4}{*}{\makecell[c]{Funcitional \\ Sites \\ Identification}} 
    & \href{https://github.com/biomed-AI/GraphBepi}{GraphBepi}~\cite{zeng2023identifying} & ESM-2 & GNN \& Bi-LSTM-based Model & LM-as-Encoder (Repr.) \\
    & \href{https://bio-web1.nscc-gz.cn/app/GPSite}{GPSite}~\cite{yuan2024genome} & ProtT5 & GNN-based Model & LM-as-Encoder (Repr.) \\
    & \href{https://github.com/fteufel/signalp-6.0}{SignalP 6.0}~\cite{teufel2022signalp} & ProtBERT & CRF Probabilistic Model & LM-as-Encoder (FM FT) \\
    & \href{https://github.com/ml4bio/USPNet}{USPNet}~\cite{shen2024unbiased} & ESM-MSA-1b / ESM-2 & Bi-LSTM-based Model & LM-as-Encoder (Repr.) \\
    \midrule
    \multirow{12}{*}{\makecell[c]{Protein- \\ Biomolecule \\ Interactions \\ Prediction}} 
    & \href{https://conplex.csail.mit.edu/}{ConPlex}~\cite{singh2023contrastive} & \gape{\makecell[c]{ProtBERT \& Morgan\\ Fingerprint Encoder}} & Cosine Distance Calculation & LM-as-Encoder (Repr.) \\
    & \href{https://github.com/Chokyotager/BIND}{BIND}~\cite{lam2024protein} & ESM-2 \& Ligand Encoder & GNN-based Model & LM-as-Encoder (Repr.) \\
    & \href{https://github.com/Luo-SynBioLab/UniKP}{UniKP}~\cite{yu2023unikp} & ProtT5 \& SMILES Transformer~\cite{honda2019smiles} & Extra Trees Model & LM-as-Encoder (Repr.) \\
    & \href{https://github.com/meyresearch/BALM}{BALM}~\cite{gorantla2024learning} & ESM-2 \& ChemBERTa-2~\cite{ahmad2022chemberta} & MLP-based Model & LM-as-Encoder (PE FT) \\
    & \href{https://github.com/WillHua127/ReactZyme}{ReactZyme}~\cite{hua2024reactzyme} & ESM-2 \& Molecule Encoder & GNN \& MLP-based Model & LM-as-Encoder (Repr.) \\
    & \href{https://github.com/wangxr0526/EasIFA}{EasIFA}~\cite{wang2024multi} & \gape{\makecell[c]{Enzyme Repr. Branch\\(w/ ESM-2 \& GearNet~\cite{zhang2023enhancing-gearnet}) \&\\ Reaction Repr. Branch}} & Cross-Attention Module & LM-as-Encoder (Repr.) \\
    & \href{https://github.com/programmablebio/saltnpeppr}{SaLT\&PepPr}~\cite{brixi2023salt} & ESM-2 & MLP Head & LM-as-Encoder (PE FT) \\
    & \href{https://github.com/microsoft/peft_proteomics}{Sledzieski et al.}~\cite{sledzieski2024democratizing} & ESM-2 & MLP Head & LM-as-Encoder (PE FT) \\
    & \href{https://github.com/MingyuJ666/ProLLM}{ProLLM}~\cite{jin2024prollm} & ProtT5 & Flan-T5-large~\cite{raffel2020exploring} & LM-as-Predictor \\
    \midrule
    \multirow{7}{*}{\makecell[c]{Multi-Task \\ Question- \\ Answering}} & \href{https://github.com/mahdip72/prot2token}{Prot2Token}~\cite{pourmirzaei2024prot2token} & ESM-2 \&  BARTSmiles~\cite{chilingaryan2022bartsmiles} & LLM Decoder & LM-as-Predictor \\
    & \href{https://github.com/UCSD-AI4H/proteinchat}{ProteinChat}~\cite{guo2023proteinchat} & GVP-GNN~\cite{jing2020learning} \& Projector & Vicuna-13B~\cite{chiang2023vicuna} & LM-as-Predictor \\
    & \href{https://github.com/hadi-abdine/Prot2Text}{Prot2Text}~\cite{abdine2024prot2text} & \gape{\makecell[c]{ESM-2, RGCN Encoder \&\\ Fusion Blocks}} & GPT-2~\cite{radford2019language} & LM-as-Predictor \\
    & ProteinGPT~\cite{xiao2024proteingpt} & ESM-2 \& GVP-GNN & LLaMA-3~\cite{dubey2024llama} & LM-as-Predictor \\
    & \href{https://github.com/xiangwenkai/MMP}{FAPM}~\cite{xiang2024fapm} & ESM-2 \& Q-Former Module~\cite{li2023blip} & Mistral-7B~\cite{jiang2023mistral} & LM-as-Predictor \\
    & \href{https://github.com/acharkq/ProtT3}{ProtT3}~\cite{liu2024prott3} & ESM-2 \& Q-Former Module & Galactica~\cite{taylor2022galactica} & LM-as-Predictor \\
    \bottomrule
    \end{tabular}
    \begin{tablenotes}
    \vspace{0.1cm}
    \item[\textbf{*}] \textbf{Repr.} - utilization of pLM-produced Representations; \textbf{FM FT} - Mull-Model Fine-Tuning; \textbf{PE FT} - Parameter-Efficient Fine-Tuning;\\ \textbf{Infer.} - Likelihood Inference
    \end{tablenotes}
    \end{threeparttable}
    \label{tab5}
\end{table*}

\subsubsection{Functional Property Prediction}

In \textbf{functional property prediction}, each protein $ x $ is mapped to a label $ y \in \mathbb{R} $ that measures a quantitative functional property, and existing studies generally employ the \textit{LM-as-encoder} scheme. For example, LMDisorder~\cite{song2023fast} utilizes the representations generated by ProtT5 to predict the disorder probability of proteins. 
On the basis of fine-tuning ESM-2, ESMtherm~\cite{chu2023protein} predicts the folding stability of proteins, PIC~\cite{kang2024comprehensive} predicts human protein essentiality, VISH-Pred~\cite{mall2024vish} is an ensemble framework for protein toxicity prediction. 
Specifically, PeptideBERT~\cite{guntuboina2023peptidebert} fine-tunes ProtBERT to predict three critical properties of peptides, i.e., solubility, hemolysis, and non-fouling. LassoESM~\cite{mi2024lassoesm} is a tailored pLM for enhanced lasso peptide property prediction, where the properties include lasso cyclase substrate tolerance, RNA polymerase inhibition activity, etc. 
Moreover, as the pH level of the reaction environment could significantly affect enzyme activity, OphPred~\cite{zaretckii2024approaching} and EpHod~\cite{gado2023deep} are recently proposed to predict the enzyme optimal pH.

Notably, the fitness of a protein is a synthetical property of how well a protein can perform its function within an organism~\cite{notin2024proteingym}. 
Experimentally, an assay of deep mutational scanning (DMS)~\cite{fowler2014deep} measures the effects of hundreds to thousands of mutants on a single protein. DMS assays are performed for different proteins with varied functions using various forms of experimental measurements~\cite{meier2021language}. Relevant influencing factors include stability, folding efficiency, binding affinity, etc. That is to say, instead of describing a specific property presented in all proteins, fitness labels assess the comprehensive function performance of a limited set of homologous proteins. Meanwhile, the effect of mutants indicates how the function of mutants changes compared to the wild-type, which is tightly connected to fitness. 

Recent studies have revealed that pLMs play a significant role in the zero- and few-shot predictions of protein fitness and mutant effects. After pre-training on massive protein sequences that survived through natural evolution, pLMs understand what sequences are plausible (i.e., probably encode proper function) while the others are invalid. Practical results demonstrate that \textit{the likelihoods inferred from pLMs correlate well with protein fitness}~\cite{meier2021language,ding2024protein,gordon2024protein}. By comparing the language modeling probabilities assigned to the mutant sequence and wild-type sequence,
ProteinGym~\cite{notin2024proteingym} calculates the mutant effect on fitness based on existing pLMs in a zero-shot manner, and Brandes et al.~\cite{brandes2023genome} have predicted all possible missense variant effects in the human genome with an ESM-1b-based workflow. 

Besides, fine-tuning pLMs on certain labeled protein fitness values leads to more robust predictions. Few-Shot Learning for Protein Fitness Prediction (FSFP)~\cite{zhou2024enhancing} is an effective training strategy designed to optimize pLMs under extreme data scarcity. By combining advanced techniques of meta-transfer learning, learning to rank, and LoRA, FSFP significantly boosts the performance of pLMs
in predicting protein fitness based on only tens of labeled single-site mutants. Then, Denoising Protein Language Models (DePLM)~\cite{wangdeplm} predicts mutation effects by refining evolutionary information captured in pLMs. Conceptually, evolutionary information could comprise both property-relevant and irrelevant information, with the latter acting as “noise” for the specific prediction task at hand. DePLM contains an ingenious denoising diffusion framework for the likelihoods produced by pLMs, effectively filtering out the irrelevant information to improve mutation effect predictions.

\subsubsection{Functional Class Annotation}

\textbf{Functional class annotation} can be acknowledged as multi-class, hierarchical, or multi-label classification problems in different situations. In annotating prokaryotic viral proteins, each viral sequence is mapped to one of nine key functional classes. Flamholz et al.~\cite{flamholz2024large} construct a pLM-based classifier, enabling fresh regions within the viral sequence space to be annotated meaningfully. In annotating enzymes, each enzyme sequence is mapped to a unique EC number. Fernandez et al.~\cite{fernandez2023exploring} explore different representation strategies of manual feature engineering, frequency space encoding, and pLM encoding, and compare various alternative classifiers in the prediction head. 
Besides, GraphEC~\cite{song2024accurately} performs EC number prediction based on geometric graph learning, where the featured graphs are constructed using the pLM-produced sequence representations and ESMFold-predicted structures. 

In Gene Ontology (GO) function prediction, each protein $ x $ is mapped to a set of labels $ \{ y_1, y_2, ..., y_n \} $ where $ y_i \in \{ 0, 1 \} $ denotes whether the $ i $-th GO term is annotated to the protein. In recent years, the NetGO series~\cite{you2019netgo,yao2021netgo,wang2023netgo} has exhibited remarkable performance in large-scale functional annotations, where pLMs have driven the latest improvement. On the basis of NetGO 2.0, the NetGO 3.0~\cite{wang2023netgo} incorporates a new model, LR-ESM, which leverages ESM-1b to present each protein and undergoes training with logistic regression. Similarly, SPROF-GO~\cite{yuan2023fast} leverages the pLM-based representation for initial prediction, and that is further advanced through a label diffusion algorithm.
Specifically, there are hierarchical inclusion relationships between GO terms, i.e., the semantic proximity between multi-class labels should be considered in the prediction. Based on representations produced by ESM-2, DeepGO-SE~\cite{kulmanov2024protein} generates multiple approximate models of GO, which predict the truth values of statements about protein functions.
The truth values over those models are aggregated to ultimately determine GO functions.

\subsubsection{Functional Site Identification}

On a micro level, proteins perform functions by some significant sites rather than the whole macromolecule. Given a protein, \textbf{functional site identification} aims to reveal which residues are of functional sites, and simultaneously identify the specific type of function if necessary. For example, 
GraphBepi~\cite{zeng2023identifying} is a graph model that predicts binding probabilities of B-cell epitope for each residue. Starting with a protein sequence, a protein graph is constructed based on the AlphaFold2-predicted structure, where pLM-learned per-token sequence representations are specifically taken as node attributes. A deep learning model consisting of GNNs and Bi-LSTM is subsequently trained for the prediction. Then GPSite~\cite{yuan2024genome} is a multi-task network aiming to predict binding residues of biomolecules (e.g., DNA, RNA, peptide, etc.) on proteins. The predicted structure from ESMFold and the representation produced by ProtT5 are similarly processed into a geometric-aware attributed protein graph. Subsequently, a shared GNN captures the common binding-relevant characteristics, and individual prediction heads are trained for ligand-specific binding site prediction. 

Besides, signal peptides (SPs) are short sequences that regulate protein secretion and translocation across all organisms, currently categorized into five types. SignalP 6.0~\cite{teufel2022signalp} combines a pLM with a conditional random field model to identify which residues belong to the SP region and simultaneously infer the SP type. Then, Unbiased Organism-agnostic Signal Peptide Network (USPNet)~\cite{shen2024unbiased} is another deep learning model proposed for SP classification and cleavage-site prediction. In USPNet, a pLM like ESM-MSA-1b or ESM-2 serves as an encoder to enrich representations of sequences, which facilitates the prediction derived by additional learning modules. 

\subsubsection{Protein-Molecule Interaction Prediction}

Proteins generally carry out functions in concert with additional molecules, which can be another protein, DNA, RNA, or drug. 
In addition to identifying the binding sites solely on proteins, predicting the protein-molecule interaction states is equally important. Given paired protein and molecule as input, \textbf{protein-molecule interaction prediction} is designed to determine whether they interact, measure relevant properties about the interaction, or precisely identify the interaction sites. For example, 
ConPLex~\cite{singh2023contrastive} is proposed to predict the interaction probability between drugs and protein targets. ConPLex aligns the embedded molecular fingerprints with the pLM-generated protein representations by contrastive learning, and calculates the distance between a pair of aligned representations for binary prediction. 
Similarly, using language models to encode proteins and ligands, BALM~\cite{gorantla2024learning} learns protein-ligand binding affinities by optimizing their cosine similarity in a shared latent space. 
Then, while taking pLM-produced representations and molecular graphs as inputs, BIND~\cite{lam2024protein} can predict the drug-target affinity values and discriminate between active and decoy ligands. 

Notably, enzymes are workhorses for various biological processes, binding to substrates and releasing products in diverse catalyzed reactions. 
In understanding the functionality of enzymes, UniKP~\cite{yu2023unikp} aims to predict enzyme kinetic parameters (i.e., enzyme turnover number, Michaelis constant, catalytic efficiency) from protein sequences and substrate structures. 
While taking AA sequences to represent enzymes and molecular graphs of substrates and products to describe catalyzed reactions, ReactZyme~\cite{hua2024reactzyme} has benchmarked the enzyme-reaction prediction, ranking enzymes by their catalytic ability for specific reactions. In addition,  EasIFA~\cite{wang2024multi} is designed to annotate enzyme active sites specific to reactions, predicting the enzymatic activity of residues by aligning the input information of proteins and enzymatic reactions.

Besides, parameter-efficient fine-tuning and prompting techniques have been extensively explored in protein-protein interaction (PPI) prediction. For example, Brixi et al.~\cite{brixi2023salt} fine-tune the final few layers of ESM-2 together with a prediction head to identify PPI sites. 
Sledzieski et al.~\cite{sledzieski2024democratizing} fine-tune ESM-2 with LoRA for binary PPI prediction, achieving state-of-the-art performance while greatly reducing memory cost. 
Recently, Protein Chain of Thought (ProCoT)~\cite{jin2024prollm} has tried to enhance the reasoning capability of LLM tailored for PPI prediction, specifically connecting the concepts of signaling pathways and natural language prompts, discovering the interaction between upstream and downstream proteins undergoing multiple biological signal transductions.

\subsubsection{Multi-Task Question-Answering} 

Despite the fantastic performance achieved by the abovementioned methods, adapting pLMs to each task over and over again can still be time-consuming and computationally expensive. 
Especially when fresh tasks come up, the consumption of resources will continue to increase in new training rounds. 
To avoid such trouble, efforts are made to create \textbf{multi-task question-answering} frameworks capable of making varied predictions within a unified model, where the \textit{LM-as-Predictor} scheme is typically employed.
For example, Prot2Token~\cite{pourmirzaei2024prot2token} possess a customized "bi-steam encoders - unified decoder" architecture, integrating a pLM (i.e., ESM-2) and a molecular LM with an autoregressive LLM to perform a variety of protein function predictions. In addition to the encoded protein and molecule sequences, a specific "task token" is also introduced to the LLM decoder to prompt what predictions should be made, which covers over ten tasks like stability prediction, fluorescence prediction, GO function prediction, human PPI prediction, etc. 

As an intuitive and inclusive data form, natural language carries the question-answering process. Given a question and related contexts as prompts, LLMs can generate textual descriptions that encompass broader knowledge rather than fixed labels. Consequently, ChatGPT-like systems are developed to enable users to upload proteins and engage in question-answering interactions to gain insights. For example, Prot2Text~\cite{abdine2024prot2text} combines GNNs and ESM-2 to encode a protein into a fused representation and employs GPT-2 to generate the protein’s text description. 
ProteinChat~\cite{guo2023proteinchat} consists of a structure encoder (i.e., GVP-GNN), a projection layer, and a general LLM (i.e., Vicuna-13B~\cite{chiang2023vicuna}), which are responsible for producing protein embedding, adapting protein to natural language, and decoding answers, respectively.
Similarly, ProteinGPT~\cite{xiao2024proteingpt} integrates protein sequence and structure encoders (i.e., ESM-2 and GVP-GNN) with linear projectors for seamless representation adaptation, connected to a general LLM (i.e., LLaMA-3) to generate logically consistent responses. 
Moreover, both FAPM~\cite{xiang2024fapm} and ProtT3~\cite{liu2024prott3} employ the Querying Transformer (Q-Former)~\cite{li2023blip} module to bridge the protein-text modalities, thereby connecting pLM encoder with LLM decoder and achieving accurate protein-to-text generations.

\subsection{Protein Design}

Over millions of years of evolution, natural proteins possess a wide range of functions, which support the running of living systems. However, existing proteins occupy only a minimal fraction of the protein space (illustrated in Figure \ref{fig0}), suggesting that enhanced or even novel functions might be encoded within the extensive unseen protein sequences. Rather than passively watch the slow process of natural evolution unfold, scientists investigate protein design to produce new proteins with desired functions. The critical challenge lies in efficiently exploring the vast protein space to find a manageable number of protein sequences that are plausible, functionally significant, and diverse. 
Depending on whether starting from existing proteins or scratch, we can classify protein design into two main categories~\cite{notin2024machine}: redesign and \textit{de novo} design.

\subsubsection{Protein Redesign}

\begin{figure}[!t]%
  \centering
  \includegraphics[width=\linewidth]{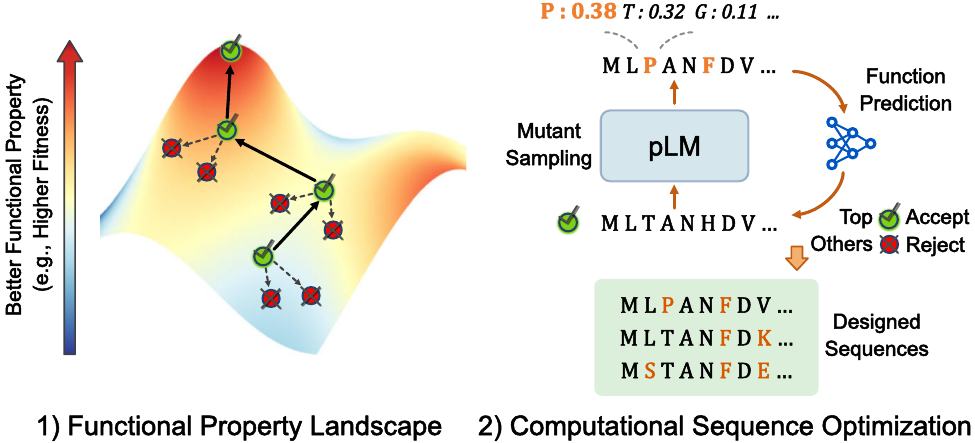}
  \captionsetup{font=small}
  \caption{\textbf{Protein Redesign: Function-Oriented Protein Sequence Optimization}. 1) Directed evolution is performed on the functional property landscape over sets of protein mutants. While the landscape exists conceptually, it is generally not fully revealed. Therefore, protein sequence optimization includes iterative rounds of mutant sampling and functional property validation. In each round, only advantaged mutants remain candidates for the next round. 2) Computational protein sequence optimization methods employ pLMs to guide mutant sampling, and leverage the predicted functional property to filter candidate sequences. 
  }
  \label{fig10}
\end{figure}

Protein redesign initializes the exploration of protein space from existing proteins, aiming to enhance existing functional properties. Directed evolution~\cite{arnold2018directed,packer2015methods} is a classic experimental method that emulates the working mechanism of natural evolution in labs, involving an iterative process that alternates between diversification and screening. In each cycle, candidate proteins experience random mutations, followed by experimental measurement of the target functional property (generally the fitness). The most favorable variants are retained as candidates, and the iterative process continues until the desired design goal is achieved~\cite{notin2024proteinnpt}. As shown in Figure \ref{fig10}-1, it is considered a sequence optimization process within the fitness landscape, aiming to step from a given seed (existing sequence) to the optima (functionally enhanced sequence). Nevertheless, the effectiveness and efficiency of this optimization process are limited in two ways: random navigation leads to lots of unnecessary mutants, and repeated measurements of functional properties impose a considerable experimental burden. Consequently, efforts are dedicated to computational sequence optimization to infer superior protein variants, aiming to lessen the laboratory workload and improve the success rate.

In recent years, pLMs have played crucial roles in computational protein redesign, aiding in simulating fitness landscapes and identifying advantageous mutations. Concretely, autoencoding pLMs have learned inter-residue relationships from large-scale pre-training. Taking input as a single wild-type protein sequence $ \mathbf{X} = \left[ x_1, x_2, ... x_L \right] $, pLMs are expert in producing conditional likelihoods, such as the 'wild-type marginal'  $ p\left( x_{i}' | \mathbf{X} \right) $ conditioned on the original sequence, or the 'masked marginal' $ p\left( x_{i}' | \mathbf{X}_{-i} \right) $ conditioned on the sequence with interested site masked. Those inferred likelihoods should probably adhere to the law of natural protein sequences. Therefore, an important assumption emerges in protein sequence optimization: \textit{mutations with high pLM-inferred likelihoods tend to be evolutionarily plausible}~\cite{hie2024efficient}.

The distribution of protein sequences learned by pLMs is primarily exploited to assist mutant sampling. Hie et al.~\cite{hie2024efficient} perform affinity maturation of human antibodies by employing pLMs to recommend plausible AA substitutions. Ensembled ESM-1b and ESM-1v models compute the 'wild-type marginal' likelihoods of all possible single-residue substitutions on the antibody variable regions, and the substitutions exceeding wild-type likelihood with a consensus are accepted for further analysis. 
This pLM-guided redesign strategy leads to impressive results, enhancing the affinity of seven antibodies by evaluating only twenty or fewer new variants of each antibody over just two rounds of laboratory evolution. 
Similarly, Johnson et al.~\cite{johnson2021generating} use Gibbs sampling from autoencoding pLMs (e.g., ESM-1b, ProtBERT) to generate novel protein sequences that retain critical characteristics of relevant natural sequences. 
The same idea holds for multiple-sequence-based pLMs, which produce conditional likelihoods based on aligned homologous sequences. Sgarbossa et al.~\cite{sgarbossa2023generative} propose an iterative protein design method that directly leverages the MLM objective to generate new sequences with ESM-MSA-1b. The resulting sequences are scored at the same level as natural sequences across the homology, co-evolution, and structure-based measures. 

In addition, the pLM-guided mutant sampling and function prediction are usually performed collaboratively to contract the range of candidate proteins across iterations. The demonstration can be found in Figure \ref{fig10}-2. 
Tran et al.~\cite{tran2023protein} utilize pLMs to pinpoint the mutation hotspots and suggest substitutions, then train a fitness prediction model to select top-performance variants. 
LLM-GA~\cite{teukam2024integrating} combines pLM with genetic algorithms to optimize enzyme feasibility and turnover dynamics: ESM-2 directs the creation of a pool of mutants, which are screened using specific fitness measurements. 
EvoOpt~\cite{yamaguchi2022evoopt} leverage ESM-MSA-1b to produce multiple mutant sequences by inferring the randomly masked positions in the original MSA. The generated sequences are subjected to zero-shot fitness prediction, and then the top-ranked proteins are preserved as the proposed candidates. 
Pro-PRIME~\cite{jiang2024general} is a temperature-guided pLM designed to suggest protein mutants with enhanced stability and activity. It is first utilized for zero-shot mutation suggestion and subsequently fine-tuned for fitness prediction. The eventual top-K mutants are adopted for further experimental analysis. 
Notably, Darmawan et al.~\cite{darmawan2023sampling} develop a \textit{in silico} evaluation framework that holistically compares pLM-based iterative protein redesign methods. Optimized protein sequences are evaluated for three core criteria: relevance, quality, and diversity, ensuring the designed proteins are functionally relevant and significant while differing from known proteins.

The fundamental goal of protein redesign is to improve the desired functional properties of existing proteins, and the aforementioned sequence optimization strategies achieve this goal by excluding low-fitness mutants. In each iteration, they sample the distribution learned by sequence-based pLMs for evolutionary plausible proteins, which are not controllably biased toward specific functional properties of interest. 
In enhancing the effectiveness and efficiency of protein redesign, efforts are made to ensure that the mutant sampling process is attuned to desired functions.
Conceptionally, function-enhanced pLMs play a role here. As shown in Figure \ref{fig6}-3, Regression Transformer~\cite{born2023regression} and ProteinNPT~\cite{notin2024proteinnpt} learn the bidirectional correlation between protein sequences and functional labels. At a specific masked position, they infer the likelihood over the amino acid vocabulary conditioned on other unmasked tokens and a given functional label. On this basis, we can steer the generation of optimized protein sequences with the property of interest. 
Moreover, EVOLVEpro~\cite{jiang2024rapid} is a few-shot active learning framework designed to improve protein activity rapidly, and its effectiveness has been demonstrated by comprehensive experiments across six therapeutically relevant proteins. 
In silico, EVOLVEpro equips ESM-2 with a regression model to learn the activity landscape for each specific protein, thereby guiding the directed evolution process. 
In each round of directed evolution, a limited number of EVOLVEpro-suggested mutants are evaluated via experimental assays. The data obtained from these experiments are subsequently used to train the model incrementally and predict mutation candidates for the next round. 
Over multiple rounds, EVOLVEpro can effectively lead scientists to new proteins with significantly improved desired properties. 

Besides, text-guided protein editing emerges as a new computational protein redesign task. Given an initial protein sequence and a prompt describing the desired functional property, we expect to get edited protein sequences that possess the potential to reach our design goal. ProteinDT~\cite{liu2023text} enables text-guided protein editing based on ProteinCLAP (illustrated in Figure \ref{fig6}-4), which aligns the representation space of protein sequence and textual description. The input protein sequence and text prompt undergo individual encoding and are fused into a latent code, which is further decoded as the optimized protein sequence. 
Similarly, ProtET~\cite{yin2024multi} comprises two stages: contrastive learning aligns the protein and biotext language model encoders, then the fused representations from original protein sequences and textual instructions serve as the condition for generating edited protein sequences. 
Moreover, ChatDrug~\cite{liu2023chatgpt} employs conversational LLMs like ChatGPT to edit drugs (e.g., small molecules, peptides, and proteins) through textual descriptions. 
Users are allowed to ask LLMs to update the drug editing results iteratively. 
In conclusion, these text-guided protein editing methods demonstrate positive results in computational evaluations, although still away from laboratory validation.

\subsubsection{\textit{De Novo} Protein Design}

Rather than engineering existing proteins, \textit{de novo} protein design aims to propose new functional proteins without reference sequences. It is highly challenging as it requires the model to grasp precisely what sequence and structure will achieve the desired function within the vast protein space. Meanwhile, there are distinct advantages of revealing functions never seen in nature and offering complete control over the design progress. 

Generally, \textit{de novo} protein design is implemented by reversing the sequence-structure-function paradigm: specify a desired function, design a structure executing this function, and find a sequence that folds into this structure~\cite{chu2024sparks}. It is widely acknowledged as two steps: \textbf{protein structure generation} and fixed-backbone sequence design (i.e., \textbf{inverse folding})~\cite{kortemme2024novo}. That is to say, the more conserved protein structures are leveraged to impose greater constraints on exploring the vast sequence space. 

\begin{itemize}[leftmargin=8pt, itemsep=0pt, topsep=1pt, parsep=1pt]
\item In terms of \textbf{protein structure generation}, diffusion models have recently achieved remarkable success~\cite{liu2023generative}. For example, RFDiffusion~\cite{watson2023novo} has demonstrated exceptional performance in generating backbone structures for topology-constrained and unconditional protein design. 
Chroma~\cite{ingraham2023illuminating} is a diffusion-based generative model tailored for proteins and protein complexes. Its generation can be guided toward diverse properties through conditioning settings. 
\item Regarding \textbf{inverse folding}, in addition to graph-based models like ProteinMPNN~\cite{dauparas2022robust}, structure-enhanced pLMs have played a significant role. They can infer reasonable protein sequences while correctly understanding the input structures. 
As illustrated in Figure \ref{fig5}-2, LM-Design~\cite{zheng2023structure} implants a structural adapter into pLM to endow it with structure awareness. As it is trained to reconstruct the corrupted protein sequence based on the provided backbone structure, LM-Design enables the refinement of protein sequences. During inference, a given structure $ \mathbf{S} $ is projected to an initial sequence $ \mathbf{X}^{(0)} $, 
and then the generated sequence is sampled following Markov process $ X^{(t)} \sim p\left( X^{(t)} | X^{(t-1)}, S \right) $ in an iterative manner until some fixed number of steps. 
Subsequently, ProseLM~\cite{ruffolo2024adapting} is another structure-conditioned protein design method based on the adaption of pLMs, where the structural context is incorporated through a set of parameter-efficient adapters. By taking ProGen2, an autoregressive pLM, as the foundation model, proseLM can generate protein sequences autoregressively for a given backbone structure. 
Similarly, InstructPLM~\cite{qiu2024instructplm} successfully aligns a frozen backbone encoder with a frozen pLM decoder by training a protein structure-sequence adapter. 
Besides, ESM-IF1~\cite{hsu2022learning} employs GVP layers to extract geometric features, followed by a generic Transformer. ESM-IF1 undergoes supervised training of structure-conditioned autoregressive sequence generation, where the training data is augmented by comprising AlphaFold2 predicted structures. 
\end{itemize}

These two steps could even be combined into an "end-to-end" model. For example, Pinal~\cite{dai2024toward} is a protein design framework that aims to generate protein sequences under the guidance of natural language, with text-to-structure and structure-to-sequence stages seamlessly integrated. The textual description of functions is first translated into structural information through an encoder-decoder network, and then protein sequences are generated based on the structure \& description through a re-trained structure-aware pLM. 

In recent years, the rise of autoregressive pLMs has introduced new ideas to protein modeling and design. It's feasible to implement \textbf{protein sequence generation} directly in unconditional or function-conditioned manners. 
As the composition of training data can significantly influence the generation distribution of AI models, ensuring alignment between the proteins that models learn and their intended applications is acknowledged as essential for improved protein sequence generation~\cite{ruffolo2024designing}. 
For example, ProGen2~\cite{nijkamp2023progen2} is primarily designed to generate diverse sequences when pre-trained on universal proteins. 
Subsequently, fine-tuning enables ProGen2 models for family-specific protein sequence generation. Besides, as shown in Figure \ref{fig6}-2, ProGen~\cite{madani2023large} and ZymCTRL~\cite{munsamy2024conditional} can propose proteins in response to specific prompt tokens. In addition to zero-shot inference, fine-tuning ProGen using curated tag-sequence instances can enhance its generation of proteins from families with sufficient homologous samples, and fine-tuning ZymCTRL enables the generation of designated proteins that have a higher probability of meeting \textit{in silico} filters and displaying activity similar to their natural counterparts.

Protein design is not completed in a one-way prediction but a complex problem encompassing multiple essential links. In practice, experts would reason with domain knowledge and organize dynamic collaborations between a series of prediction or experimental tools. In addition to working as an independent model, general LLMs like GPT-4 have driven the development of multi-agent systems, bringing a new perspective in promoting protein design. ProtAgent~\cite{ghafarollahi2024protagents} is an LLM-based platform proposed for \textit{de novo} protein design, where multiple AI agents cooperate within a dynamic environment while each possesses the capability of knowledge retrieval, protein modeling, physics simulation, or results analysis, respectively. ProtAgent demonstrates the potential to minimize the requirement for human interference during the iterative problem-solving process.

\section{Applications}

In the field of protein science, the impact of Protein Language Models (pLMs) is exhibited in not only \textit{in silico} modeling but also wet-lab recognized applications. In this section, we discuss some significant application topics, i.e., antibody design, enzyme design, and drug discovery. 
In each application topic, we concisely show the problem's background, summarize the role of LLMs within a general workflow, and present certain studies with solid experimental analysis. 

\subsection{Antibody Design}

\begin{figure*}[!t]%
  \centering
  \includegraphics[width=0.8\linewidth]{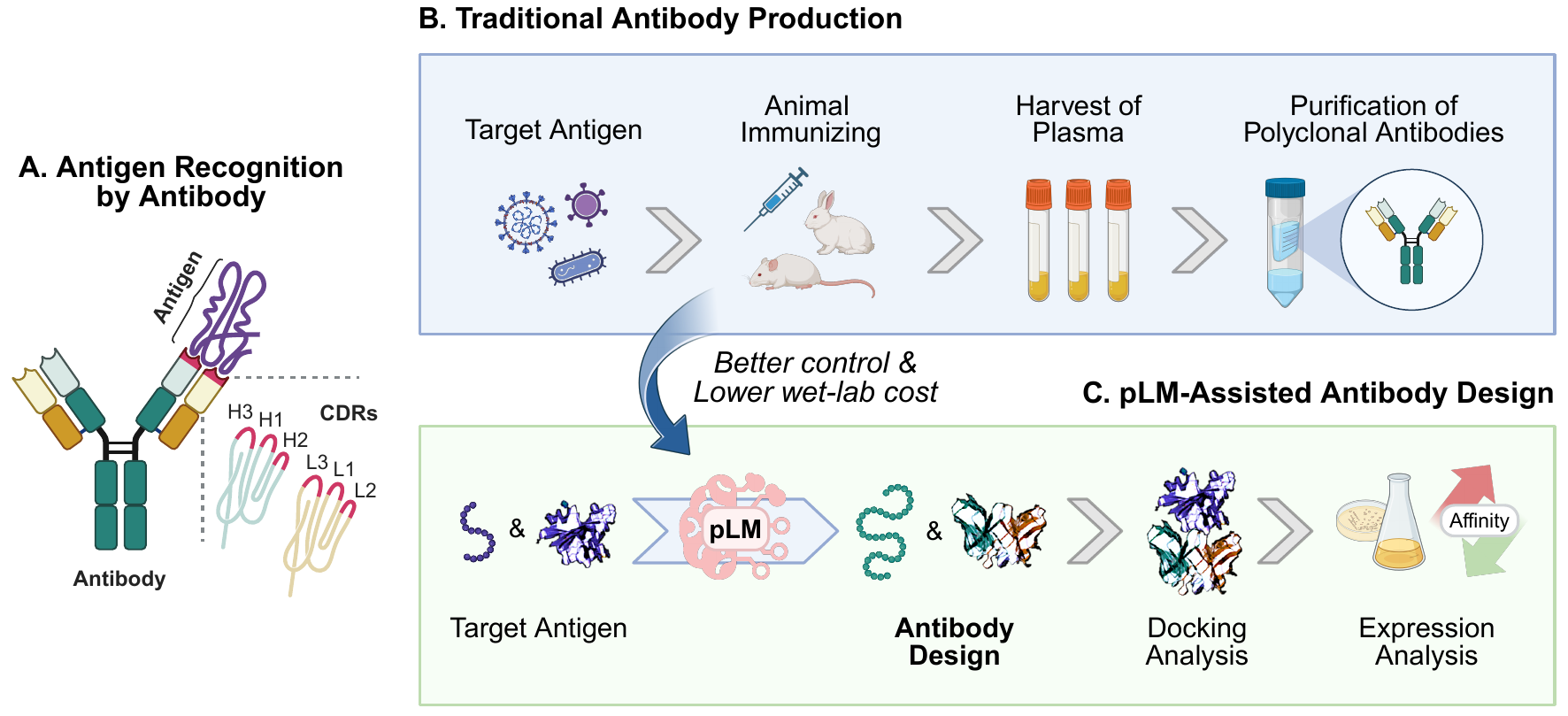}
  \captionsetup{font=small}
  \caption{Overview of \textbf{Antibody Design}. (A) In the process of recognizing antigens, the complementarity-determining regions (CDRs) of antibodies are critically important. (B) Traditional antibody production is limited by the immunity of animals. (C) Protein language models can be adopted to propose antibody sequences and structures. The visualization of antibody and antigen structures is derived from PDB entry 7T72, which features SARS-CoV-2 neutralizing antibodies. This figure is created in BioRender.com.} 
  \label{fig11}
\end{figure*}

Antibody is a type of protein that exists in the immune system, aiding in body defense by identifying and neutralizing harmful entities like bacteria and viruses, also known as antigens~\cite{rajewsky1996clonal}. 
As illustrated in \ref{fig11}-A, central to the recognition process is the interaction between antigen and the complementarity-determining regions (CDRs) of antibody. Then, with antigens neutralized, antibodies successfully help maintain the normal functions of living organisms.
Once a life body has defeated an antigen, the antibodies synthesized would remain in the bloodstream, offering protection if the same antigen appears again in the future. 

Figure \ref{fig11}-B shows the overall workflow of traditional antibody production. The target antigen is typically introduced into an animal, such as a rabbit or a mouse, and their natural immune response can generate antibodies in the blood. Then, the desired antibodies are obtained with the plasma collected and purified~\cite{laustsen2021animal}. Despite the massive practical success, antibody production like this is restricted to the innate immunity of animals. On the one hand, it's difficult to control the quality of produced antibodies. On the other hand, wet lab experiments are usually cost-extensively and time-consuming. To overcome these problems, massive efforts are dedicated to artificial antibody design, where pLMs have played an important role in recent years.

As shown in Figure \ref{fig11}-C, pLM-based models can assist antibody design by proposing antibody sequences and structures that specifically bind to the target antigen. These designed antibodies are then tested in computational docking and experimental expression analyses. Their biological effectiveness, especially the antigen-binding affinity, is verified by the expression results in yeast cells. For example, 
PALM-H3~\cite{he2024novo} is proposed for the \textit{de novo} generation of antibody heavy chain CDR 3 (CDR-H3) that meets the desired binding specificity. 
In evaluations, PALM-H3 can generate antibodies that not only target the SARS-CoV-2 wild-type but also adapt to its emerging variants, including Alpha, Delta, and XBB variants. Besides, Shanker et al.~\cite{shanker2024unsupervised} demonstrate that a structure-informed pLM, i.e., ESM-IF can be used to guide the evolution of antibodies. 
Thirty variants of two therapeutic clinical antibodies are screened for their effectiveness in treating severe acute respiratory SARS-CoV-2 infection.
In response to antibody-escaped viral variants of concern BQ.1.1 and XBB.1.5, the designed antibodies have demonstrated notable performance, showing up to a 25-fold enhancement in neutralization and a 37-fold improvement in affinity.

\subsection{Enzyme Design}

\begin{figure*}[!t]%
  \centering
  \includegraphics[width=0.80\linewidth]{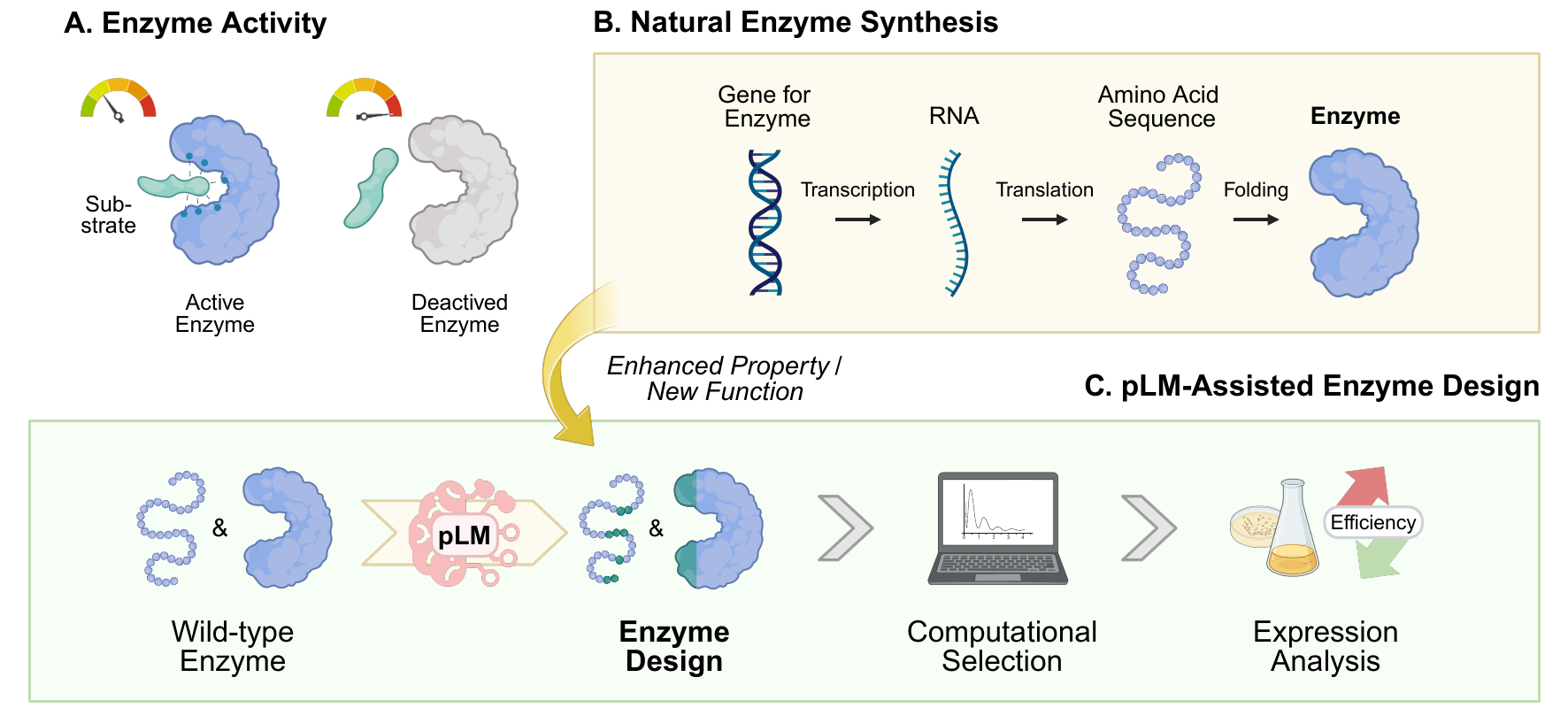}
  \captionsetup{font=small}
  \caption{Overview of \textbf{Enzyme Design}. 
  (A) Natural enzymes are usually sensitive to the reaction environment.
  (B) Genes determine the function of natural enzymes. 
  (C) pLMs can assist in the design of enhanced enzymes.
  This figure is created in BioRender.com.} 
  \label{fig12}
\end{figure*}

Enzymes are valuable proteins that act as natural molecular optimization machines to speed up chemical reactions. As illustrated in Figure \ref{fig12}-A, in a suitable environment, enzymes interact with substrates at their active sites, enabling a broad range of biological activities~\cite{robinson2015enzymes}. For example, enzymes break down large molecules like glucose into smaller ones that the body can use for energy~\cite{romero2015dissecting} and assist in DNA replication~\cite{saffhill1973effects}. 
Figure \ref{fig12}-B shows the process of natural enzyme synthesis. Wild-type enzymes are ultimately determined by the information recorded in genes and are produced through the transcription, translation, and folding flows by life bodies. However, wild-type enzymes usually have short lifespans and low stability~\cite{craik2011proteases}, making it difficult to meet the emerging modern requirements. Therefore, scientists are dedicated to designing enzymes with enhanced properties (e.g., thermostability) and even new catalytic functions, which could help in tackling major global challenges, such as the shortage of energy, pollution of the environment, and lack of sufficient food~\cite{bornscheuer2012engineering}.

Figure \ref{fig12}-C shows a typical workflow of pLM-assisted enzyme design. Scientists can utilize pLMs to optimize wild-type enzymes for desired functions, and the designed enzymes possess the potential to exhibit significant efficiency throughout computational selection and further expression analysis. 
For example, InstructPLM~\cite{qiu2024instructplm} employs pLMs to generate optimized enzyme sequences based on specific backbone structures. In wet lab experiments (the expression analysis in E. coli), it is demonstrated that the redesigned enzymes of PETase and L-MDH exhibit efficacy levels that exceed those of their wild-type.
Besides, Johnson et al.~\cite{johnson2024computational} conduct computational scoring and experimental assessment of enzymes generated by three contrasting models.
With the help of a novel computational framework named COMPASS, part of the generated sequences with high scores are selected for the further \textit{in vitro} assays. Experimental results demonstrate that certain enzyme sequences designed by pLMs could express in E. coli and show activity.

\subsection{Drug Discovery}

In cells, biological functions are usually not performed by individual proteins but rather by their dynamic interactions with each other or with molecules, and the temporary complexes are specifically formed. These interactions are essential for managing the cellular activities essential for life, such as transcription, splicing, etc~\cite{gebauer2004molecular,cozzolino2021protein}. Viewed from the pathogenesis perspective, the interactions between drugs (mostly small molecules) and their targets (generally proteins) can effectively regulate the health of organisms. Understanding the target protein and revealing drug-target interaction is critical in drug discovery~\cite{hou2022discovering}, which could raise great medical significance and economic value. 

\begin{figure*}[!t]%
  \centering
  \includegraphics[width=0.80\linewidth]{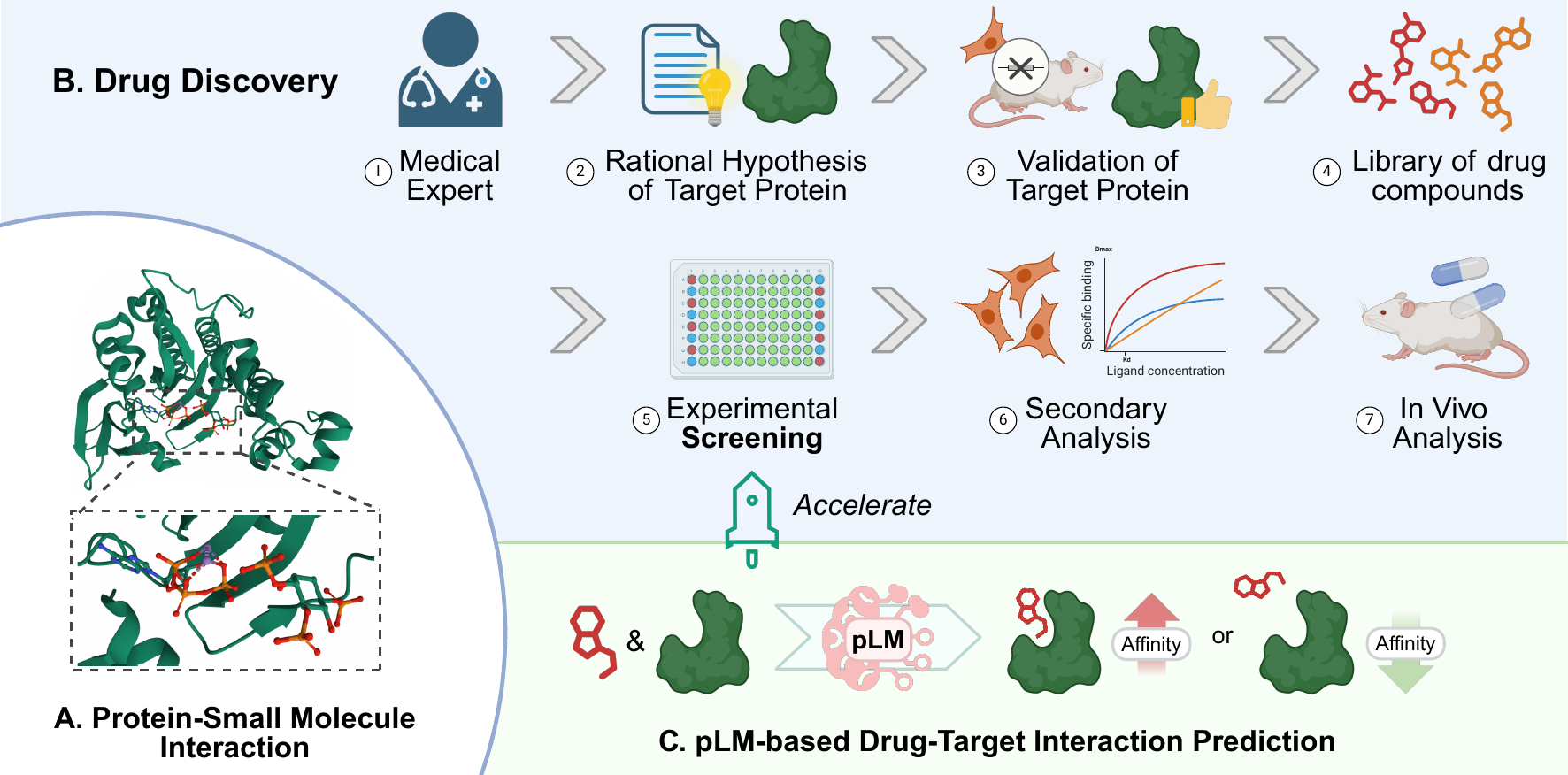}
  \captionsetup{font=small}
  \caption{Overview of \textbf{Drug Discovery}. (A) A visualization for protein-ligand interaction, which is obtained from PDB entry 8PPB. (B) Overview of a general drug discovery process~\cite{hughes2011principles}. (C) pLMs are employed to predict drug-target interactions, potentially accelerating drug discovery by aiding in drug candidate screening. This figure is created in BioRender.com.} 
  \label{fig13}
\end{figure*}

Figure \ref{fig13}-B outlines a general process of drug discovery. Initially, often in academic settings, scientists would hypothesize that activating or inhibiting a target protein could have therapeutic effects in a disease state. The next step is to confirm if the target exactly plays a crucial role in the onset and progression of diseases. Subsequently, compound screening takes place, aiming to select drugs that can interact with the specific target from a vast library of drug compounds. This process typically involves high-throughput experiments, which require substantial time and materials~\cite{hughes2011principles}. 
After the preliminary screening, secondary analyses are conducted to assess the selected compounds in various biological models and experiments, ensuring their efficacy and safety. Finally, these compounds are tested in animal models to evaluate their pharmacological activity within a biological system.

For decades, researchers have attempted to accelerate drug discovery by computational strategies. Notably, pLM-based models demonstrate the potential to complement screening by predicting the interactions between drugs and target proteins. 
For example, TransDTI~\cite{kalakoti2022transdti} is an effective computational workflow that employs pLMs to categorize drug-protein target interactions as active, inactive, or intermediate. With prediction results confirmed through molecular docking and simulation analysis, TransDTI can accurately identify different drug interactions with two proteins, i.e., MAP2k and TGF$\beta$. 
In addition, ConPLex~\cite{singh2023contrastive} investigates contrastive learning in drug-target latent space. The usability of ConPlex is demonstrated by testing 19 kinase-drug interactions, with 12 interactions confirmed. Specifically, four of these interactions demonstrated exceptionally high binding affinity, including a potent EPHB1 inhibitor with a dissociation constant of 1.3 nM when interacting with the compound PD-166326.

\section{Future Directions}

In this survey, we have thoroughly reviewed the progress in the interdisciplinary field of computational protein science and large language models, illustrating that protein language models possess a foundational understanding of proteins and can promote the advancement in essential protein modeling problems, i.e., structure prediction, function prediction, and protein design. Despite the achievements, there are still several challenges in this field, which also indicate the future directions.

\subsection{Data Scarcity}

Breakthrough AI for protein studies are primarily driven by abundant protein sequence and structure data. Gratefully, excellent scientists have invested countless efforts in laboratory protein sequencing and structure solutions and have generously shared their findings in public databases. Moreover, AlphaFold has predicted extensive protein structures that are highly reliable. However, data scarcity is still a tough issue in many specific practical tasks. For example, the imbalanced species representations in protein sequence databases lead to consistent species bias that can be detrimental for protein design applications~\cite{ding2024protein};
and it's difficult to assess the robustness of models developed for protein fitness prediction and redesign as lacking large-scale benchmarks with consistent ground-truth labels~\cite{notin2024proteingym}. The enhancement of multi-modal learning is also limited by the scale of protein-text datasets developed with expertise~\cite{pei2024leveraging}. To address the challenge of data scarcity, it's promising to expand and diversify the training data through augmentation or synthesis methods~\cite{zhou2025decoding}, and empower AI models with the capability to learn from limited instances.

\subsection{Protein Interaction Modeling}

In living organisms, proteins usually do not exist in a simple form of "single-chain sequence, monomer structure, \& independent function". Instead, multiple polypeptide chains could fold together to form protein complexes, 
and various protein molecules can interact physically within specific biomolecular contexts. 
These protein-protein interactions are crucial for regulating a wide array of biological activities. 
Despite the current success of language model techniques in computational protein science, extending protein language models (pLMs) to effectively model protein interactions remains a significant challenge, which involves foundational tasks such as understanding, structure prediction, and function prediction of protein complexes. 
One of the primary reasons is the limited availability of data on protein interactions. 
For instance, the Protein Data Bank (PDB)~\cite{wwpdb2019protein} contains relatively few protein complexes, and the Observed Antibody Space (OAS)~\cite{kovaltsuk2018observed} has limited paired variable heavy and light chain data. 
Meanwhile, most existing protein interaction data may not have been fully exploited in training pLMs. For instance, data pipelines, such as those employed by AlphaFold2~\cite{jumper2021highly} and ESMFold~\cite{lin2023evolutionary}, often treat polypeptide chains individually, even when they are components of larger protein complexes.
Navigating the challenge alongside the opportunity, several efforts have been made to advance protein interaction modeling.
For example, IgBert and IgT5~\cite{kenlay2024large} are antibody-specific language models designed to consistently handle the paired and unpaired variable region sequences. 
Linker-Tuning~\cite{zou2023linker} extends ESMFold, 
a method originally designed for single-chain structure prediction,
to predict heterodimer structures. 
Furthermore, RDE~\cite{luo2023rotamer}  focuses on understanding the effect of mutations on protein-protein interactions. 
In the future, we believe that there will be a surge in research and novel findings aimed at addressing this formidable challenge. 
Continued efforts in this field hold great promise for enhancing our understanding of complex biological systems and advancing computational protein science.

\subsection{Explainability}

As summarized in Sections 3 and 4, pLMs extract statistical patterns within natural protein sequences, and then pLM-based structure prediction, function prediction, and protein design methods capture information flows within the sequence-structure-function paradigm. However, the learned principles are latent in hundreds of millions of model parameters. Although AI models such as ESM-2 and AlphaFold2 have been widely acknowledged and applied in scientific exploration, scientists have barely gained explicit physical insights from them~\cite{doerr2024protein}. Experts would take "black-box" prediction models as tools that output probably accurate "observations", while do not directly get aids in the understanding of new theories. If we can distill the knowledge learned by AI models in a form humans can readily understand, protein sciences could be significantly promoted in a new manner. 
These days, there have been individual studies on this. Zhang et al.~\cite{zhang2024protein} demonstrate that pLMs learn evolutionary statistics of interacting sequence motifs. Then, \href{https://interprot.com/}{InterPro}~\cite{simon2024interplm} is proposed to extract, analyze, and visualize human-interpretable latent features from pLMs. 
In the future, the more in-depth Explainable AI (XAI)~\cite{xiong2024towards} research for protein science still presents a challenging but influential blue ocean.

\subsection{Bridging Computational and Experimental Research} 

Scientific discovery is a multifaceted process that involves multiple interconnected stages covering hypothesis, experiments, and data~\cite{wang2023scientific}. To date, protein language models are mainly accelerating protein science research with dry lab predictions. Biologists still undertake heavy workloads like rational reasoning with domain knowledge, planning the combination of computational tools \& experimental facilities, and performing wet lab experiments. It is acknowledged that general LLMs like GPT-4~\cite{achiam2023gpt} have exceptional abilities in solving complex problems by empowering multiple intelligence agents, and that could be utilized in driving end-to-end scientific research. Then, there have been some surprising results in the field of chemistry. Artificial intelligence systems empowered by GPT-4, like ChemCrow~\cite{m2024augmenting} and Coscientist~\cite{boiko2023autonomous}, can autonomously design, plan, and perform complex experiments by incorporating multiple scientific research tools. In a similar way, LLMs should hold the potential to drive further research in protein science~\cite{ghafarollahi2024protagents}. In the future, we look forward to more mature studies in autonomous research that bridge the gap between computational and experimental protein science, thereby helping biologists in all stages of work and accelerating real-world scientific discovery.

\subsection{Computational Efficiency}

In the recent progress of AI, scaling law~\cite{kaplan2020scaling} stands out as a particularly instructive concept. As LLMs scale up in parameters, data, and computing, larger models acquire unprecedented emergent capabilities and demonstrate substantially improved performance. It is observed the development of pLMs follows this trend as well. For example, ESM-3~\cite{hayes2025simulating} is scaled up to 98 billion parameters, and xTrimoPGLM is trained at a massive scale of 100 billion parameters. No matter for big companies or academic groups, the high computational cost poses a tough issue, and the computational efficiency of pLMs should be improved. 
Therefore, it's in demand to understand the scaling behaviors for protein language modeling and formulate optimized computational schemes. 
Recently, there have been certain ongoing studies questioning whether pLMs are compute-optimal and reached the unanimous conclusion of non-optimal~\cite{serrano2024protein,cheng2024training}, exploring how to scale down pLMs for better efficiency while maintaining their expressiveness~\cite{fournier2024protein,vieira2024scaling}, and implementing novel techniques like FlashAttention~\cite{dao2022flashattention} to achieve efficient inference, training, and fine-tuning of pLMs~\cite{celik2024efficient,faesm2024}. 
In addition to these preliminary explorations, in the future, more efforts could be dedicated to implementing optimal models in the constraint of predetermined computation budgets.

\section{Conclusion}

We present a comprehensive survey of computational protein science in the era of LLMs, containing broad content from background concepts to the latest advancements. 
First, we outline the biological basis and data profiles in protein modeling. 
Second, we review three categories of pLMs with abilities to comprehend amino acid sequences, recognize structural and functional information, and bridge multiple biomedical languages. 
Next, we introduce the utilization and adaptation of pLMs, highlighting their significant impacts on structure prediction, function prediction, and protein design.
Then, we specify the application potentials of pLMs in antibody design, enzyme design, and drug target discovery. 
Finally, we share the promising future directions in this fast-growing field. 
% Our goal is to offer a systematic and comprehensive review for readers with either AI or biology backgrounds, presenting research focus and breakthroughs in computational protein science, and motivating further explorations. 

\ifCLASSOPTIONcaptionsoff
  \newpage
\fi

\balance
\bibliographystyle{unsrt}  
\bibliography{main}  

\end{document}